\documentclass{revtex4}
\usepackage{graphics}
\usepackage{epsfig}
\usepackage{colordvi} 
\begin{document}

\newcommand{\RS}{\RawSienna}
\newcommand{\MG}{\Magenta}
\newcommand{\BL}{\Blue}

\newcommand{\nl}{\nonumber \\ }
\newcommand{\bq}{ \begin {equation} }
\newcommand{\eq}{\end{equation}}
\newcommand{\ba}{\begin{eqnarray}}
\newcommand {\ea} {\end {eqnarray} }

\newcommand{\be}{\begin{equation}}
\newcommand{\ee}{\end{equation}}
\newcommand{\nn}{\nonumber}
\newcommand{\bea}{\begin{eqnarray}}
\newcommand{\eea}{\end{eqnarray}}

\def \litwo {{\rm{Li_2}}}
\def \litr {{\rm{Li_3}}}
\def \lifo {{\rm{Li_4}}}
\def \ep   {\epsilon}

{\tt 
\noindent
DESY 04-222
\\
SFB-CPP-04-61
\\
hep-ph/0412164}

\bigskip

\title{%
Master integrals for massive two-loop Bhabha scattering in QED
}
\author{M. Czakon}
\affiliation{Institut f\"ur Theoretische Physik und Astrophysik, Universti\"at W\"urzburg, 
Am Hubland, D-97074 W\"urzburg, Germany}
\affiliation{Institute  of Physics, University of
Silesia, Uniwersytecka 4, PL-40-007 Katowice, Poland }
\email{gluza@us.edu.pl,Tord.Riemann@desy.de,mczakon@yahoo.com}
\author{J. Gluza}
\affiliation{DESY, Platanenallee 6, 15738 Zeuthen, Germany}
\affiliation{Institute  of Physics, University of
Silesia, Uniwersytecka 4, PL-40-007 Katowice, Poland }
\author{T. Riemann}
\affiliation{DESY, Platanenallee 6, 15738 Zeuthen, Germany$\,^{*}$}
\begin{abstract}
We present a set of
scalar master integrals (MIs) needed for a complete
treatment of 
massive two-loop corrections to Bhabha scattering in QED,
including integrals with arbitrary fermionic loops.
The status of analytical solutions for the MIs is reviewed and
examples of some methods to solve MIs analytically are worked
out in more detail. 
Analytical results for the pole terms  in $\epsilon$ 
of so far unknown box MIs with five internal lines are given.
\end{abstract}


\maketitle


\section{\label{s1}Introduction}
Bhabha scattering is proposed to be measured at the International
Linear Collider (ILC) in the very forward direction with a precision
that would allow to determine the luminosity with an accuracy of
10$^{-4}$
\cite{Aguilar-Saavedra:2001rg,Hawkings:1999ac,Lohmann:2004nn,Lohmann:2004nn2}.
The Monte Carlo programs which were in use for the analysis of LEP
data aimed at a slightly lower accuracy, and one could neglect
certain two-loop corrections there; for a discussion see \cite{Jadach:2003zr}. 
At the ILC, the theoretical prediction of the differential cross
section for Bhabha scattering
\ba
\label{eq1}
e^+(p_2) + e^-(p_1) \to e^+(p_4) +  e^-(p_3) 
\ea
has to include the complete virtual photonic two-loop
corrections. 
This is a highly nontrivial task, but with substantial recent
progress in several directions.
Here, we will concentrate mainly on efforts to determine the two-loop
corrections in a calculational scheme with a finite electron mass $m$,
and with a regularization of both UV and IR divergences with
dimensional regularization
\footnote{Alternatively, one may assume both
  massless photons and electrons from the beginning \cite{Bern:2000ie}.
}.
Because the Monte Carlo programs for the treatment of real
bremsstrahlung assume a finite photon mass at intermediate
states of the calculation
 \cite{Jadach:1996is,Melles:1997qa,Arbuzov:1995qd,Arbuzov:1996jj}, one
will have to care about this fact when results will be finally combined. 
This might be done similarly as in 
\cite{Glover:2001ev,Mastrolia:2003yz}.
Alternatively, the soft photon
bremsstrahlung can be recalculated completely in $D$ dimensions as in
\cite{Bonciani:2004qt}, where 
this is done for the simplest  subset of the corrections. 

In the QED model with three leptonic flavors one has 154
Feynman diagrams (with all the 1PI diagrams, but without loops in
external lines), among them 68 double-box diagrams \cite{web-masters:2004nn}.
Besides the usual problems of efficient bookkeeping, the main problem
is the evaluation of the loop integrals.
One has to solve Feynman integrals with $L\leq 2$ loops and $N\leq 7$
internal lines,  
\bea
\label{bh-48}
G(X) &=& \frac{1}{\left(i\pi^{d/2}\right)^L} \int \frac{d^Dk_1 \ldots d^Dk_L~~X}
     {(q_1^2-m_1^2)^{\nu_1} \ldots (q_j^2-m_j^2)^{\nu_j} \ldots
       (q_N^2-m_N^2)^{\nu_N}  }  ,
\eea
where $X = 1, k_{1\alpha}, k_{1\alpha}k_{2\beta}, \ldots $ stands for
tensors in the loop momenta.
This might be done by
a procedure with three subsequent steps:
\begin{itemize} 
\item[(i)] reduce all tensorial loop integrals to scalar integrals,
\item[(ii)] reduce these
to a smaller set of scalar master integrals (MIs),
\item[(iii)] evaluate the MIs.
\end{itemize}
A completely numerical approach might also be possible \cite{Passarino:2004nn}.
For checks in the Euclidean region this has been proven to be a
powerful tool
\cite{Binoth:2000ps,Binoth:2003ak}; see Appendix \ref{sec-a2}.
 
Step (i) {may be} considered to be solved by now, 
the second one 
is solved for Bhabha scattering with this article 
{\footnote{In \cite{Czakon:2004tg}
a complete set of prototypes has been shown, for one fermion flavor,
but without reference to the exact definition of the MIs.}}, 
and step (iii) is solved
for all self energies and vertices, but remains largely unsolved for
the two-loop boxes.
So, the evaluation of the two-loop boxes  is the remaining bottleneck.
Indeed, very few of the 33 double-box master integrals have been
determined 
completely
\cite{Smirnov:2001cm,Bonciani:2003cj,Heinrich:2004iq,Czakon:2004tg} or
to some extent 
\cite{Heinrich:2004iq}.

The article is organized as follows. 
In Section \ref{sec-proto}, the 
diagrams and prototypes for two-loop Bhabha scattering
are identified  and the method for their evaluation 
in terms of MIs
is outlined.
We discuss in Section \ref{sec-master} the complete set of MIs and
give an overview with figures and tables.
The status of analytical (or semi-analytical) solutions of the known MIs
is summarized. 
Some typical techniques for MI evaluation are demonstrated.
The section includes also a discussion of MIs with numerators.
We close with a  Summary. 
Appendix \ref{sec-a2} includes an 
extension of the numerical method for the evaluation of Feynman
diagrams with sequential sector decompositions,
which allows to treat integrals with irreducible numerators.
The final  Appendix \ref{a-hpl} defines a complete and compact set
of Harmonic Polylogarithms up to weight 4.

\section{\label{sec-proto}From diagrams to prototypes and master integrals}
\begin{figure}[b]
\epsfig{file=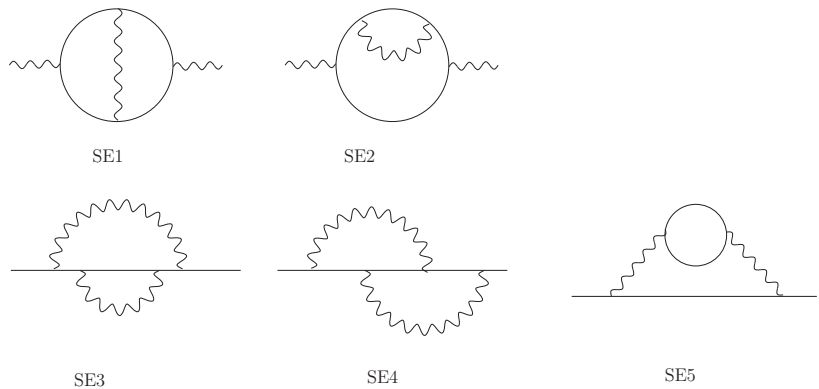, width=11cm}
\caption{The two-loop self-energy diagrams for massive Bhabha
  scattering.
SE3 to SE5 are needed for the renormalization of external fermion lines.
}
\label{SEprot}
\end{figure}

\begin{figure}[htbp]
\epsfig{file=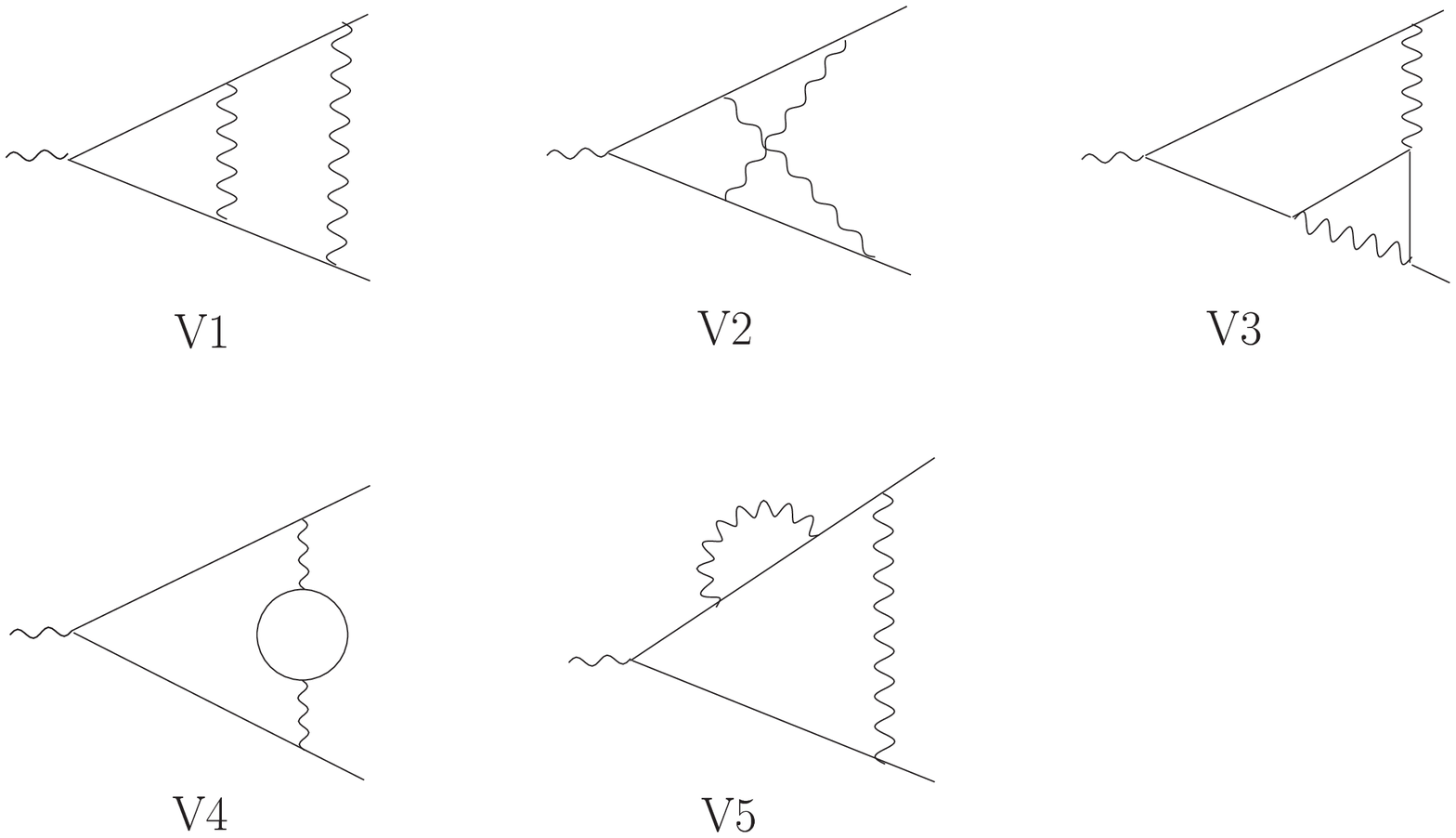, width=8cm}
\caption{The two-loop vertex diagrams for  massive Bhabha scattering.}
\label{beta}
\end{figure}

\begin{figure}[ht]
\epsfig{file=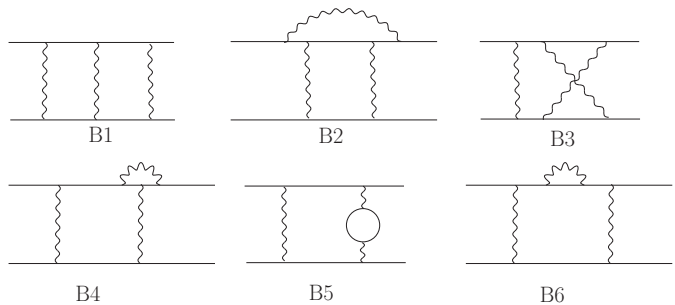, width=9cm}
\caption{The two-loop box diagrams for massive Bhabha scattering.}
\label{LL2}
\end{figure}

In a theory with only electrons and photons, one has in the
't Hooft-Feynman gauge 52 1PI two-loop Feynman diagrams,
all of the double-box type.
After adding to this number the 1P-reducible diagrams without
loop insertions in external lines, there are 94 diagrams.
We will call this the one-flavor case \footnote{Bhabha scattering
  with only electrons and photons, we call it here the {\em one-flavor case},
  should not be mixed with what is often called the $N_f=1$ QED Bhabha
  scattering. The latter includes only the two-loop diagrams SE1, SE4,
  V4, B5, 
  and for the soft photon treatment one has to take into account only
  B5.
}.
Due to the existence of up to two closed fermion loops in certain
diagrams, a complete picture of the process arises when besides the
electron two additional flavors are taken into account.
Then, there are 68 double boxes and 154 two-loop Feynman diagrams in
total \footnote{We take into account that due to Furry's theorem some
  diagrams vanish pairwise.}. 
Fortunately, there are much less two-loop structures to be calculated;
they are represented by prototypes.
Prototypes are irreducible (sub)diagrams of a certain topology with
account of the various, different masses of internal lines.
All the Feynman integrals $G(X)$ (see  (\ref{bh-48})) with the same
topology and propagators, but with arbitrary powers $\nu_i>0$ 
of these propagators, and potentially with irreducible numerators $X$
correspond to one prototype.  
In Figures \ref{SEprot} to \ref{LL2} 
we show the sets of two-loop self-energy, vertex
and box diagrams for which the scalar MIs are
needed.
As mentioned, there are two bosonic self energies SE1 and SE2, and the other
three fermionic self energies renormalize the external lines. 
Further, there are five two-loop vertex and six two-loop box diagrams 
for Bhabha scattering.

Technically,   
one has to calculate all the Feynman integrals $G(X)$ related to these
diagrams by a reduction of integrals with irreducible
numerators $X$ and denominators with higher powers $\nu_i$ to a smaller
set of scalar master integrals.

We have determined such a set of scalar MIs for the virtual
two-loop corrections to Bhabha scattering with the Laporta-Remiddi
algorithm \cite{Laporta:1996mq,Laporta:2001dd}.
Realizations of the algorithm are {\tt SOLVE}
  \cite{Remiddi:2004nn}, the {\tt Maple} program {\tt AIR}
  \cite{Anastasiou:2004vj}, and the {\tt C++} library {\tt
    DiaGen/IdSolver} \cite{Czakon:2004uu2}.
We use  
{\tt DiaGen/IdSolver} \footnote{%
The package {\tt DiaGen}  has already
been used in several other projects  
\cite{Czakon:2002wm,Awramik:2002wn}, 
, while {\tt IdSolver} is a new package by
  M.C \cite{Czakon:2004uu2}.
Both have been used recently also for a four loop project \cite{Czakon:2004bu}.
Furthermore, we are using Fermat \cite{Lewis:2004nn}, FORM
\cite{Vermaseren:2000nd}, Maple, and Mathematica.
}%
, which allows to tackle two problems: 
(i) derive an appropriate set of algebraic equations with integration by
parts (IBP) \cite{Chetyrkin:1981qh} 
and Lorentz invariance (LI) identities \cite{Gehrmann:1999as}
\footnote{The  Lorentz invariance identities have been useful for
  algorithmic optimization but did not reduce the number of MIs.}, and
(ii) determine a list of master integrals by solving this set of equations.  
The procedure is heuristic.
To be safe about the completeness of the solution, one has to solve
the system of equations to high powers of numerators and 
denominators. 
If the coefficients are kept as exact functions of the kinematical
variables $s, t, m^2$ (and for several flavors of a further scale  $M^2$), the
algorithm is time and computer memory consuming.  
To minimize computer ressources usage, evaluation
homomorphisms are used, {\em i.e.} the system is solved by projecting
the coefficients to the field of rational numbers with suitably chosen
values of the parameters. 
Let us mention that, from the point of view of complexity, the  
complete two-loop Bhabha scattering calculation with {\em massless} fermions is as
complicated as the two-loop massive {\em vertex} case: 
on a 2 GHz Pentium PC with 1 GB memory,
it takes minutes to be solved.
Further, the number of MIs is moderate, {\em e.g.} 5 massless MIs
compared to 22  MIs for the massive box diagram B3.

There is some arbitrariness in the choice of masters.
We prefer to present a set of MI without numerators.
But we then have to allow for higher powers of propagators ($\nu_i >
1$, dotted lines).  
This choice has one basic advantage: the MIs are independent
of the momenta flowing inside loops. 
Of course, in the set of solutions for MIs, there are algebraic
relations between scalar integrals with numerators and those with
dotted denominators: so one always has a freedom of choice.
For an explicit evaluation of the MIs, this is of importance; see the
discussions in the subsequent sections. 

In Table \ref{table} we give a list of the net numbers of
(two-loop)+(one-loop) master integrals
needed for the evaluation of all two-loop vertex and box diagrams for
the one-flavor case.
The case of several flavors is separately discussed in Section \ref{sse-4}.

\section{\label{sec-master}Master Integrals}
Here we will describe the sets of master
integrals needed for the calculation of all  Feynman
integrals for the  diagrams in Figures \ref{SEprot}  to \ref{LL2}.   
We use a nomenclature where {\em e.g.}
{\tt V3l2m} is the name of the integral for a {\em Vertex} with  {\em 3
  lines}, among them {\em 2 massive} lines;
{\tt B5l4md} is the name of the integral for a {\em Box} with {\em 5 lines},
among them {\em 4 massive} lines and one 
line with a {\em dot}.
Sometimes there are several candidates for the same name, {\em e.g.} 
{\tt V6l4m1} and {\tt V6l4m2} in Figure \ref{beta4a}, or {\tt
  V4l1m1d1} and {\tt V4l1m1d2} in 
Figure \ref{beta5a}.

Because they are lengthy, we give in this article practically no explicit 
expressions for the master integrals.
Instead,
for all the self energies and vertices, 
they may be found in \cite{web-masters:2004nn}
in form of a human readable
Mathematica file, {\tt MastersBhabha.m}. 
We determined all these MIs, and several of the box masters, in terms of
Harmonic Polylogarithms  (HPLs)
\cite{Remiddi:1999ew}.
The file allows a determination of the expressions in form of
polylogarithms.
We used HPLs up to the weight 4 and give in Appendix \ref{a-hpl} a complete
basis for their systematic calculation.
The complete list of HPLs may be found in the Mathematica file {\tt
  HPL4.m} in   \cite{web-masters:2004nn}.  
Numerical checks were performed with the numerical integration package
{\tt sectors.m}, see Appendix \ref{sec-a2}.

We evaluated most of the MIs with the method of differential
equations \cite{Kotikov:1991hm,Remiddi:1997ny},
which has been described in detail in many papers, {\em e.g.}
\cite{Laporta:2001dd,Bonciani:2003cj,Bonciani:2003hc,Bonciani:2003te},
and we will 
not repeat this here.  Nevertheless, it might be useful to indicate some technical details of our calculations.
The master integrals are defined in a Minkowskian metric, and 
the external momenta are introduced in (\ref{eq1}).
With $(p_1+p_2)=(p_3+p_4)$ and $p_i^2=m^2$, we define
$s= (p_1+p_2)^2,\;t=(p_1-p_3)^2,\;u=(p_1-p_4)^2$.
The analytical results are expressed by dimensionless variables
$x$ and $y$, 
\begin{eqnarray}
\label{defx}
x = \frac{\sqrt{-s+4}-\sqrt{-s}}{\sqrt{-s+4}+\sqrt{-s}},
\end{eqnarray}
corresponding to $s=-(1-x)^2/x$, and $y$ is obtained by replacing $s$
by $t$.    
We set the electron mass to unity, $m^2=1$.

In the rest of the article, we discuss the set of master
integrals
and their expansions in the parameter $\epsilon = (4-D)/2$. 
First we treat only electrons and photons.
The diagrams with additional flavors and thus with a second mass
scale are discussed in Section \ref{sse-4}.

\subsection{One-loop Master Integrals}
\begin{figure}[thb]
\epsfig{file=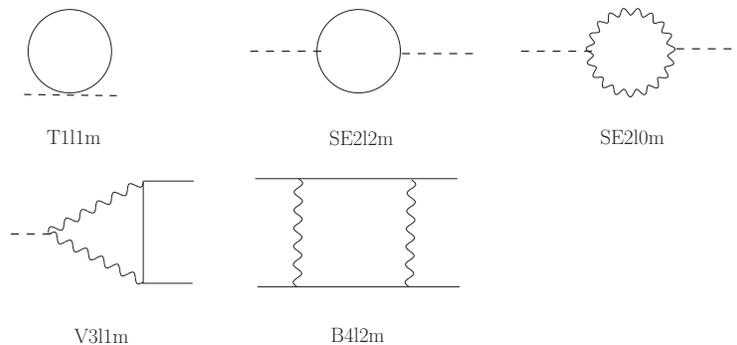, width=10cm}
\caption{The five one-loop MIs. External solid (dashed) lines describe 
on- (off-)shell momenta.
}
\label{beta1}
\end{figure}

There are five one-loop master integrals needed for the evaluation of the
two-loop diagrams; see Figure \ref{beta1}.
Our normalization of the momentum integrals is chosen such that the
one-loop tadpole becomes:
\begin{eqnarray}
{\tt T1l1m} 
&=& 
 \frac{     e^{\epsilon\gamma_E}}{i\pi^{D/2}} 
\int \frac{d^{D}q}{q^2-1} = \frac{1}{\epsilon} +1
+\left(1+ \frac{\zeta_2}{2}\right)  \epsilon 
+\left(1+ \frac{\zeta_2}{2} - \frac{\zeta_3}{3}\right)  \epsilon^2 
+\ldots
\label{norm}
\end{eqnarray}

For completeness, we should mention that a full calculation of the Bhabha
scattering process (\ref{eq1}) also includes  the  one loop
corrections in the electroweak
Standard Model (plus some higher order corrections).
For their treatment  we refer to
\cite{Consoli:1979xw,Bohm:1984yt,Bohm:1986fg,Bardin:1990xe,%
Bardin:1997xq,Beenakker:1997fi,Bardin:1999yd,Kobel:2000aw,%
Gluza:2004tq,Lorca:2004dk,Lorca:2004fg}. 

\subsection{\label{sec-2l2p}Two-loop 2-point Master Integrals}
\begin{figure}[htbp]
\epsfig{file=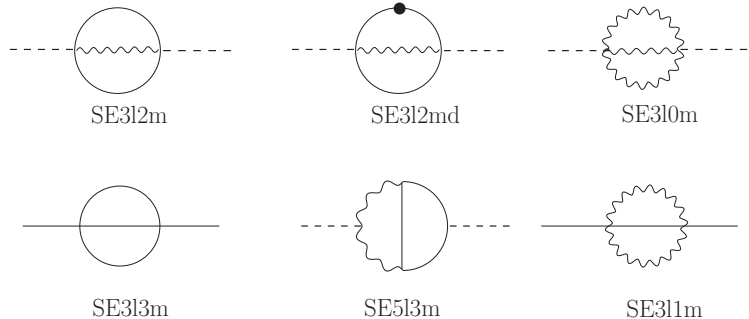, width=10cm}
\caption{The six two-loop 2-point MIs. External solid (dashed) lines describe 
on (off) -shell momenta.
}
\label{beta2}
\end{figure}

There are six two-loop 2-point MIs; see Figure \ref{beta2}.
The MI  {\tt SE3l0m} 
may be expressed by two subsequent one-loop momentum integrations
 in terms of $\Gamma$ functions whose $\epsilon$-expansion is trivial
 \cite{'tHooft:1972fi,Surguladze:1989ez}.    
Explicit expressions may also be found in \cite{Bonciani:2003hc},
quoted from \cite{Bonciani:2001}.
The MIs {\tt SE3l1m} and {\tt SE3l3m} for the renormalization of external
fermion legs are needed only on the mass shell; they may be 
calculated with {\tt ON-SHELL2} \cite{Fleischer:1999tu}, with an
appropriate  change of normalization with respect to (\ref{norm}).
For MI  {\tt SE3l3m} see also \cite{Broadhurst:1990ei,Broadhurst:1991fi}. 
The MI  {\tt SE5l3m} was first calculated to ${\cal O}(1)$ in
\cite{Broadhurst:1990ei} \footnote{The function ${\tilde I}_3$ in equation (45)
  of \cite{Broadhurst:1990ei} equals the MI $(s \cdot{\tt
    SE5l3m}[0,x])$ given in the list of masters at
  \cite{web-masters:2004nn}.} (see also \cite{Fleischer:1998nb})  
and to   ${\cal O}(\epsilon)$ in \cite{Davydychev:2003mv,Bonciani:2003hc}. 
The two remaining MIs  {\tt SE3l2m} = {\tt SE3l2m}(1,1,1,0) and  {\tt
  SE3l2md} = {\tt SE3l2m}(1,1,2,0) are defined by:
\ba
\label{2lp}
{\tt SE3l2m} (a,b,c,d)  &=&= 
- \frac{e^{2\epsilon\gamma_E}}{\pi^{D}} 
\int
\frac{d^Dk_1d^Dk_2~~(k_1k_2)^{-d}}
{[(k_1+k_2-p)^2-m^2]^{b}~[k_1^2]^{a}~[k_2^2-m^2]^{c}}.
\ea
These two MIs have been expressed in \cite{Fleischer:1999hp}
in terms of {\tt T1l1m} and the functions  {\tt SE3l2m}(1,2,2,0) and
{\tt SE3l2m}(1,1,3,0), and the
$\epsilon$-expansion is determined there  (after equation (15)) in
terms of polylogarithms.

An equivalent result, in terms of HPLs, is given in
\cite{Bonciani:2003te}, where  the integrals {\tt SE3l2m} and  {\tt
  SE3l2mN} =  {\tt SE3l2m}(1,1,1,--1) have been chosen as masters.
By an algebraic relation, valid for $m^2=1$ and $p^2=s$,
\ba
\label{2lv}
{\tt SE3l2md} &=& \frac{-(1+s)+\epsilon(2+s)}{s-4}  ~{\tt SE3l2m} +
\frac{2(1-\epsilon)}{s-4}\left( {\tt T1l1m}^2 +3~ {\tt SE3l2mN}
\right),
\ea
one may derive then {\tt SE3l2md}.

For a direct determination of the MIs with differential equations,
we constructed a differential operator for the self energies 
using the  scaling property: 
\begin{equation}
\label{2la}
{\tt SE}( \lambda p^2, \lambda m^2) = \lambda^{\text{dim}[{\tt
    SE}(p^2,m^2)]} ~{\tt SE}(p^2,m^2),
\end{equation}
where $\text{dim}({\tt SE})$ is the dimension of the 2-point MI ${\tt SE}$.
The differential operator is:
\begin{equation}
\label{2lb}
s \frac{\partial}{\partial s} {\tt SE}(s,m^2) = -m^2 \frac{\partial} 
{\partial m^2}
  {\tt SE}(s,m^2) + \text{dim}[{\tt SE}(s,m^2)]~ {\tt SE}(s,m^2).
\end{equation}
With this operator, one may derive
coupled differential equations for the
two masters and solve them with account of boundary conditions at the
kinematical point  $s=0$.
Because  the integral {\tt  SE3l2md} is one of our masters,
we reproduce it here explicitely: 
\ba
\label{2lq}
{\tt  SE3l2md}(x) &=& 
\frac{1}{2\epsilon^2} + \frac{1}{2\epsilon} 
-
\left(
\frac{1 - \zeta_2}{2}
+
 \frac{1 + x}{1- x}H[0, x]
+
  \frac{1 + x^2}{(1- x)^2}  H[0, 0, x] \right) 
+ {\cal O}(\epsilon).
\ea

In Tables \ref{2loopmast3N1} and \ref{tab-vertbox}, we list for each
of the diagrams the MIs needed for their evaluation.
Masters denoted by an asterisk are of one-loop type.

\subsection{\label{sub-vertex}Two-loop 3-point Master Integrals}
\begin{figure}[bhtp]
\epsfig{file=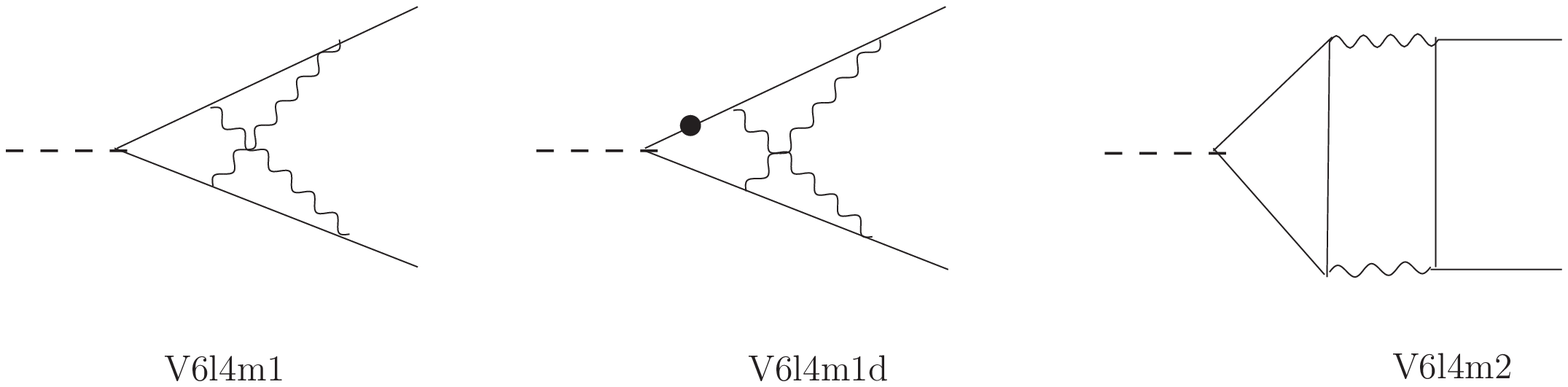, width=12cm}
\caption{The three two-loop vertex MIs with six internal lines.}
\label{beta4a}
\end{figure}

\begin{figure}[bhtp]
\epsfig{file=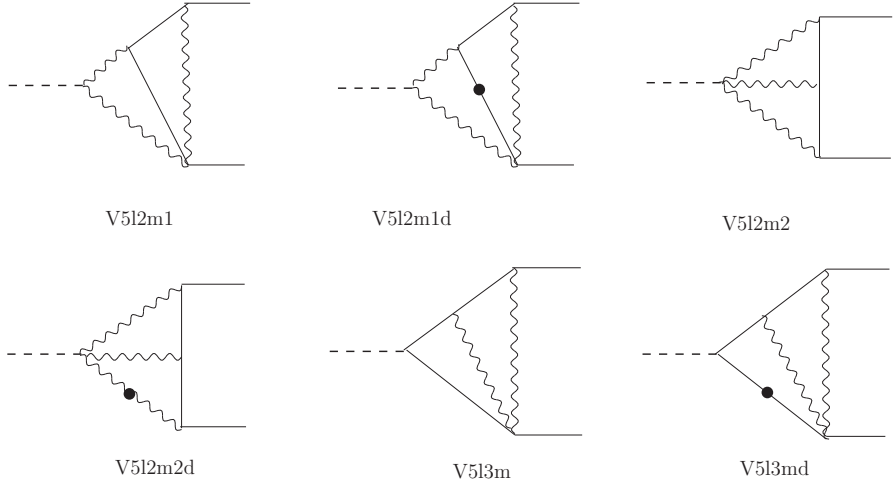, width=12cm}
\caption{The six two-loop vertex MIs with five internal lines.}
\label{beta4}
\end{figure}

\begin{figure}[hbtp]
\epsfig{file=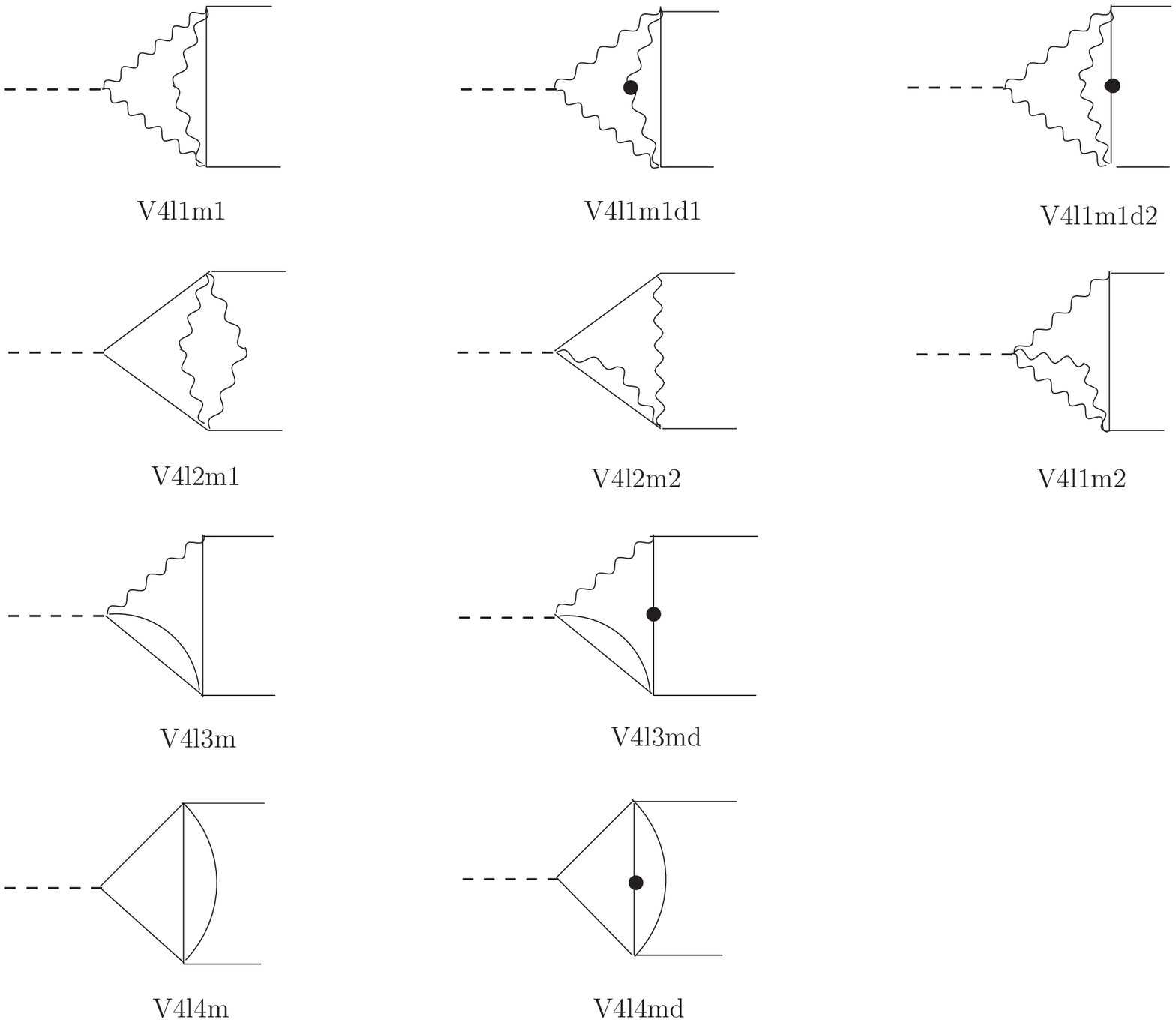, width=12cm}
\caption{The ten two-loop vertex MIs with four internal lines.}
\label{beta5a}
\end{figure}

There are nineteen two-loop 3-point MIs for Bhabha scattering.
They are shown in Figures \ref{beta4a} to \ref{beta5a}.
The equations for vertex MIs may be  obtained with a
differential operator: 
\begin{eqnarray}
s \frac{\partial }{\partial s}  = \frac{1}{4-s} \left[
2  p_2^\mu \frac{\partial}{\partial p_1^\mu} +(2-s) p_1^\mu
\frac{\partial}{\partial p_1^\mu}  \right] .
\end{eqnarray}
Another operator can be obtained by the exchange $p_1 \leftrightarrow p_2$.

In Table \ref{tab-vertbox} we list all the vertex MIs for
the various diagrams of Figure \ref{beta}.
The MIs for the evaluation of the QED (and QCD) vertex diagrams
are worked out in \cite{Bonciani:2003te,Bonciani:2003hc}.
The four  MIs of prototypes {\tt V5l2m1} and {\tt V5l2m2} are additionally 
needed for the evaluation of box diagrams B1 to B4 and have been
determined in   \cite{Czakon:2004tg}. 
It was possible to determine all the vertex MIs with the method of
differential equations.
When dotted masters (or those with irreducible numerators) are involved
one has to treat a system of coupled first order linear equations.
For prototype  {\tt V4l1m1}, three masters had to be treated together.   

As an instructive example for the use of coupled differential
equations we will study 
here the prototype {\tt V5l2m2}.
For few selected examples, we applied also alternative methods of
calculation in order to have some cross checks; see also \cite{Czakon:2004wu}.
Two of them will also be exemplified here.
One method (for MI {\tt V4l1m2}) implies the integration over an UV
divergent subloop 
and a subsequent subtraction in order to isolate an additional UV  
singularity, and the other one  (for MI {\tt V5l2m2}) introduces a
subtracted dispersion relation.  

\subsubsection{\label{a1-3}
The master integrals for the prototype {\tt V5l2m2}}  
The prototype 
\ba
\label{betav5}
   {\tt V5l2m2}(a,b,c,d,e;h,i) 
&=&
- \frac{e^{2\epsilon\gamma_E}}{\pi^{D}} 
\int
\frac{d^Dk_1d^Dk_2 
~(p_1k_2)^{-h}  ~(p_2k_2)^{-i}} 
{ 
[(k_1+k_2)^2-m^2]^{a}
[k_2^2-m^2]^{b}
[(p_1-k_1-k_2)^2]^{c} 
[k_1^2]^{d}
[(k_2-p_3)^2]^{e}
} 
\ea
has the masters  {\tt V5l2m2} = {\tt V5l2m2}(1,1,1,1,1;0,0) and  {\tt
  V5l2m2d} = {\tt V5l2m2}(1,1,1,1,2;0,0), together with 
simpler ones.
They contribute to the box diagrams B1 (planar
double box) and B3 (nonplanar double box), see Table \ref{tab-vertbox}. 
The momenta are, with the conventions used here,  in the
$t$ channel. 
Applying the corresponding differential operator (\ref{sop}), in the
$t$ channel, we wrote the
differential equations for the MI  {\tt V5l2m2} and for
{\tt V5l2m2N} = {\tt V5l2m2}(1,1,1,1,1;--1,0).
One gets:
\ba
\label{tdtV5l2m20}
t\frac{\partial }{\partial t}{\tt V5l2m2}
&=&
\frac{(-4 + s + 2t)}{2(-4 + s + t)}
\Bigl(    {\tt V4l1m2}[1, 1, 1, 2; 0] -  {\tt  V5l2m2} +2 ~{\tt
    V5l2m2d}
\Bigr)
\nl &&
 -~\frac{t}{(-4 + s + t)}~{\tt  V5l2m2}  [1, 1, 1, 1, 2; 0,-1]
-~{\tt  V5l2m2}  [1, 1, 1, 1, 2; -1,0] ,
\\ 
\label{tdtV5l2m2N0}
t\frac{\partial}{\partial t} {\tt V5l2m2N}
&=&
\frac{(-4 + s + 2t)}{4(-4+s+t)}
\Bigl(2 ~{\tt V4l1m2} [1, 1, 1, 2;-1] -(-2+t) ~{\tt V4l1m2} [1, 1, 1, 2;0]
\nl
&&+~4~{\tt V5l2m2N}\Bigr)
+\frac{t}{(-4+s+t)} ~{\tt V5l2m2} [1, 1, 1, 1, 2;-1, -1]
- {\tt V5l2m2}[1, 1, 1, 1, 2; -2, 0] .
\ea
Additional functions at the right hand side are of prototype {\tt V4l1m2}:
\ba
   {\tt V4l1m2}(a,b,c,d;g) 
&=&
- \frac{e^{2\epsilon\gamma_E}}{\pi^{D}} 
\int
\frac{d^Dk_1d^Dk_2 
~(p_1k_2)^{-g} 
}{ 
[(p_3+k_1+k_2)^2-m^2]^{a}
[p_1-p_3-k_1-k_2]^{b}
[k_1^2]^{c}
[k_2^2]^{d}
}. 
\ea
Again, we write the basic master  {\tt V4l1m2} (no dots, no
numerators) without arguments.

Besides MIs, at the right hand side of (\ref{tdtV5l2m20}) and
(\ref{tdtV5l2m2N0}) one meets 
additional scalar 
integrals, which have to be expressed by algebraic equations through
MIs:
\ba
\label{tdtV5l2m2Ny}
 {\tt V4l1m2} [1, 1, 1, 2;0] 
&=&
\frac{(1 -5 \epsilon + 6\epsilon^2) }{2 \epsilon t}
\left(4 ~{\tt V4l1m2}
+ \frac{2 - 3\epsilon}{-1 + 4\epsilon} {\tt SE3l1m}
\right) ,
\\
\label{tdtV5l2m2Nu}
 {\tt V4l1m2} [1, 1, 1, 2;-1]
&=&
\frac{2 - 3\epsilon}{4t}\left(-2 ~{\tt SE3l0m}
+
t ~{\tt SE3l1m}\right)
+
\frac{-1 + 3\epsilon}{2}
 ~{\tt V4l1m2} ,
\\
\label{tdtV5l2m2Nb}
  {\tt V5l2m2}[1, 1, 1, 1, 2; -2, 0] 
&=&
-\frac{(-1 + \epsilon)^2(-2 + 4\epsilon(1 + \epsilon) +t(1-3\epsilon)
  )}
{8(1 -
2\epsilon)^2\epsilon}
 {\tt T1l1m}^2
+
\frac{\ep(1 + 2\ep - t)}{-1 + 2\ep} {\tt V5l2m2}
\nl&&
+~
[(8- 2t-t^2)
+\ep(- 56+18t+6t^2)
+\ep^2(128 -54t-11t^2)
+\ep^3(- 96   + 60 t   + 2 t^2)]
\nl&&\times~
\frac{(-2 + 3\ep)
}{16(-1 + 2\ep)(-1 + 4\ep)}
 {\tt SE3l1m}
-
\frac{(-2\ep + 4\ep^2 - t)(-2 + t)}{4\ep(-1 + 2\ep)}
 {\tt V5l2m2d}
\nl&&+
\frac{(-2 + 3\ep)(-2 + \ep(8- 4t) + \ep^2 (-4 + t)  + t)}{8\ep t}
 {\tt SE3l0m}
\nl&&
+~
\frac{8 - 6t + t^2 - 2\ep(-4 + t)(-5 + 2t) + \ep^2(48 + -20t + t^2)}{8t}
 {\tt V4l1m2}
\ea
Three further expressions may be left out here
\cite{web-masters:2004nn}, but we have to give that for {\tt V5l2m2N}:
\ba
\label{dV5l2m2N}
{\tt V5l2m2N}
&=&
\frac{4}{1 - 2\ep}\left( \ep~{\tt V5l2m2} \right)
+\frac{t(1 + 2\ep)}{2\ep(-1 + 2\ep)}
{\tt V5l2m2d}
\nl&&
-\frac{(-2 + 3\ep)(3 -17\ep + 26\ep^2)}{8\ep(-1 +2\ep)(-1 + 4\ep) }
{\tt SE3l1m}
+
\frac{(-1 + \ep)^2(-1 + 4\ep)}{4(1 - 2\ep)^2\ep}
{\tt T1l1m}^2
\nl&&
- \frac{2 + 9(-1 + \ep)\ep}{4\ep^2 t}
{\tt SE3l0m}
+\frac{-1 + \ep}{4\ep}
{\tt V4l1m2} .
\ea
This equation may be inverted in order to eliminate   {\tt V5l2m2d} at
the right 
hand side of (\ref{tdtV5l2m20}) and (\ref{tdtV5l2m2N0}) in favor of
{\tt V5l2m2N}, 
 {\tt V5l2m2}, and simpler MIs:
\ba
\label{tdtV5l2m2N}
t\frac{\partial}{\partial t} {\tt V5l2m2N}
&=&
 -2 \left(\ep~{\tt V5l2m2}\right)
- \frac{(-1 + \epsilon)^2}{2(-1 + 2\epsilon)}{\tt T1l1m}^2
+ \frac{2 -7 \epsilon +  6\epsilon^2}{2(-1 + 2\epsilon)(-4 + t)}
\left({\tt SE3l1m}-{\tt SE3l0m}\right)
\nl
&&
-~\frac{(6 - 16\epsilon - 2t +
  5\epsilon t)}{2(-4 + t)}
{\tt V4l1m2}
,
\\ 
\label{tdtV5l2m2}
t\frac{\partial }{\partial t}{\tt V5l2m2}
&=&
 \frac{2(1+ 2\epsilon) -t}{-4 + t}{\tt V5l2m2}
+ \frac{-1 + 2\epsilon}{-4 + t}\left(2~ {\tt V5l2m2N}+{\tt V4l1m2} \right)
\nl &&
+~
\frac{2 -7 \epsilon + 6\epsilon^2}{2(-1 + 2\epsilon)(-4 + t)}
{\tt SE3l1m}
-   \frac{(-1 + \epsilon)^2}{(-1 + 2\epsilon)(-4 + t) }
{\tt T1l1m}^2 .
\ea
The dependence on the unknowns in these two equations is such that
they decouple as power series in $\ep$, and their solution is
simplified thereby.
Applying the inverted (\ref{dV5l2m2N}),  the two masters are determined
\cite{Czakon:2004tg}: 
\begin{eqnarray}
{\tt V5l2m2} 
&=&  
\frac{2x}{(-1 + x^2)}
\left( 8 \zeta_4 
+ 2\zeta_2H[0, 0, x] 
- 4\zeta_2H[0, 1, x] + 
  H[0, 0, 0, 0, x]
 + 2H[0, 0, 0, 1, x] \right)
+ \cal{O}(\epsilon) ,
\\
\label{beta6a}
{\tt V5l2m2d} 
&=&  
 -  \frac{x}{4\epsilon^2(1-x)^2}     
+ \frac{x(1 - H[0, x] -   2 H[1, x])}{2\epsilon(1-x)^2}
\nonumber \\
&&+~\frac{x}{4(1 - x)^2(1 + x)}
\bigl[ -4 - 15\zeta_2 - 4x  
+ \zeta_2x 
+ 4(1 + x)\left( H[0, x] + 2H[1, x] \right)  
\nonumber \\
&&-~2(x+3) \left( H[0, 0, x] +2 H[0, 1, x]  \right)
- 8(1+x) \left( H[1, 0, x]  +2H[1, 1, x] \right) \bigr]
+ \cal{O}(\epsilon) .
\end{eqnarray}

Another interesting opportunity appears when a system has some undotted
and dotted masters being both UV finite, but IR 
divergent, and at the same time a related integral with numerator is
both UV and IR finite
\cite{Czakon:2004wu}, 
as it is the case here.
Then, one may determine singularities in $\ep$ from the 
algebraic relations between the functions; here from  (\ref{dV5l2m2N}).   
This way one also gets the divergent parts of (\ref{beta6a}).
This may serve as a cross check here.
The dot  in  {\tt V5l2m2d} introduces overlapping IR singularities
and produces the $1/\epsilon^2$ term.

\subsubsection{\label{a1-1}Example: The master integral {\tt V4l1m2} }
The MI {\tt V4l1m2} (see Figure \ref{beta5a})  has a massless
UV divergent subloop:
\ba
\label{V4l1m2}
{\tt V4l1m2} &=& 
-\frac{e^{2\epsilon\gamma_E}}{\pi^D}
\int
\frac{d^Dk_1d^Dk_2}{[k_2^2][(k_1+k_2-p_1)^2][k_1^2-1][(k_1+p_2)^2]} .
\ea
We will use a subtraction procedure in order to isolate the  remaining UV  singularity.
The two momentum integrations may be performed subsequently:
\ba
\label{V4l1m20}
\int\frac{d^Dk_2}{[k_2^2][(k_2+k_1-p_1)^2]} &=&
i\pi^{D/2} \frac{\Gamma(1-\epsilon)^2\Gamma(\epsilon) }{\Gamma(2-2\epsilon)}
\frac{1}{[(k_1-p_1)^2]^\epsilon}    ,
\\
\label{V4l1m2x}
\int\frac{d^Dk_1}{[(k_1-p_1)^2]^\epsilon[k_1^2-1][(k_1+p_2)^2] } &=&
i\pi^{D/2} 
\frac{\epsilon(1+\epsilon)\Gamma\left(2\epsilon\right)}{\Gamma(2+\epsilon)}
~I_{\rm div},
\\
\label{V4l1m2b}
I_{\rm div} &=& \int_0^1
\frac{dxdy~x^{-1+\epsilon}(1-x)^{1-2\epsilon}}{([(1-x)(1-y)^2
  -xys]^2)^{2\epsilon}}.
\ea
The Feynman parameter integral $I_{\rm div}$  has a singularity at
$x=0$ and may be regulated by a subtraction:
\begin{eqnarray}
\label{V4l1m2c}
I_{\rm div} &=&
  \int_0^1 dx
  x^{-1+\epsilon}(1-x)^{1-2\epsilon}
\int_0^1 dy \left\{ \left[ 
f(x,y)^{-2\epsilon} - f(0,y)^{-2\epsilon}
\right]+   f(0,y)^{-2\epsilon}\right\} 
\nl
&=&
\frac{\Gamma(\epsilon)\Gamma(2-2\epsilon)}{(1-4\epsilon)\Gamma(2-\epsilon)}
 + I_{reg}
,
\\
\label{V4l1m2d}
 I_{reg} &=& 
 \int_0^1 dx~(1-x)\left[x(1-x)^2\right]^{\epsilon}\int_0^1 dy 
\frac{f(x,y)^{-2\epsilon}-f(0,y)^{-2\epsilon}}{x} ,
\ea
with
\ba
\label{V4l1m2e}
 f(x,y)=(1-x)(1-y)^2-xys.
\ea
The remaining  integrations in $I_{reg}$ are regular and can be performed
analytically or numerically after the $\epsilon$-expansion:
\ba
\label{V4l1m2f}
 I_{reg} &=& 
\int_0^1 dx (1-x) e^{\epsilon \ln[x/(1-x)^2]}
\int_0^1 \frac{dy}{x} \ln\left(\frac{f(x,y)}{f(0,y)}\right) 
\sum_{n=1}^{\infty} 
\frac{(-2\epsilon)^n}{n!} 
\left[ \sum_{k=0}^n \ln^{n-k-1} f(x,y) \ln^{k} f(0,y) \right] .
\ea
The first terms of the series expansion in $\epsilon$ for {\tt
  V4l1m2} are (see (\ref{defx})): 
\ba
 {\tt V4l1m2} &=&
\frac{1}{2\epsilon^2} + \frac{5}{2\epsilon} + 
 \frac{19}{2} -  \frac{3-13x}{2(1+x)}\zeta_2
-  \frac{1-x}{2(1+x)}\left[ \ln^2 (x) + 4 {\litwo}(x)  
\right] + {\cal O}(\epsilon).
\ea
They
coincide with our results for the MI {\tt V4l1m2} given in
\cite{Czakon:2004tg,web-masters:2004nn}.
They 
are related, {\em e.g.}, to the same MI given in \cite{Bonciani:2003hc} by the
following relation: 
${\tt V4l1m2} = 16 [e^{\epsilon\gamma_E} \Gamma(1+\epsilon)]^2 
[F_6^{-2}/(-2\epsilon)^2 + F_6^{-1}/(-2\epsilon) + F_6^{0} + \ldots ]$.

\subsubsection{\label{a1-2}Example: The master integral {\tt V4l3md} }
The master integral  {\tt V4l3md} is shown in Figure
\ref{beta5a}.
It is defined as follows:
\ba
{\tt V4l3md} 
&=& 
- \frac{e^{2\epsilon\gamma_E}}{\pi^D}\int \frac{d^Dk_1d^Dk_2}
{[k_2^2-m^2][\{k_2-(k_1+p_1)\}^2-m^2][k_1^2-m^2]^2[(k_1-p_2)^2]} ,
\ea
The MI is UV and IR divergent, and one may first integrate over the
UV divergent subloop and then treat the IR singularity by a
subtraction.
The first step gives \footnote{We use here the {\tt LoopTools}
  notations \cite{Hahn:1998yk,Hahn:2001xx}.
Our normalizations deviate corresponding to (\ref{norm}) by an additional
factor $\exp(\epsilon\gamma_E)$ and by setting $4\pi\mu^2 = 1$.
}:   
\ba
{\tt V4l3md}   
&=& 
 \frac{e^{\epsilon\gamma_E}}{i\pi^{D/2}}
\int\frac{d^Dk_1} {[k_1^2-m^2]^2[(k_1-p_2)^2]}
B_0\left[(p_1+k_1)^2;m^2,m^2 \right]      . 
\ea
Due to the dotted photon propagator, 
the integral over $k_1$ produces for $k_1 \to p_2$ an infrared
singularity.
This singularity may be isolated by a subtraction:
\ba
{\tt V4l3md} 
&=& 
  B_0\left(s;m^2,m^2\right)   B_0\left(m^2;\underline{m}^2,0\right)
+ \frac{e^{\epsilon\gamma_E}}{i\pi^{D/2}} \int \frac{d^Dk_1
  \left\{B_0\left[(p_1+k_1)^2;m^2,m^2 \right] - 
  B_0\left[s;m^2,m^2 \right] \right\} }
{[k_1^2-m^2]^2[(k_1-p_2)^2]} .
\nl
\ea
Here, the on-shell $B_0$ function with underlined argument has a
dotted line. 
The remaining integral over $k_1$  is
finite and one may evaluate it by use of a dispersion relation for the
subtracted $B_0$:
\ba
B_0\left[(p_1+k_1)^2;m^2,m^2\right] - B_0\left(s;m^2,m^2\right) 
&=& 
\frac{1}{\pi} \int_{4m^2}^{\infty} d\sigma ~\text{Im} B_0(\sigma,m^2,m^2) \left[ 
\frac{1}{\sigma-(p_1+k_1)^2}
- \frac{1}{\sigma-s} 
\right]
,
\ea
with
$\text{Im} B_0(s, m^2, m^2) = \pi \sqrt{1-4m^2/s}$.
After interchanging the order of integrations, one gets in the
argument of the $\sigma$-integration two
dotted, infrared-divergent one loop functions $C_0$ and $B_0$.
In sum, the calculation of a two-loop vertex master
integral has been reduced to the
determination of one loop functions plus dispersion integrals:
\ba
\label{bh-48d}
{\tt V4l3md} 
&=& 
 B_0\left(m^2;\underline{m}^2,0\right) B_0\left(s;m^2,m^2\right)   
 - 
 \int_{4m^2}^{\infty}
 \frac{d\sigma}{\pi}  
\text{Im} B_0(\sigma,m^2,m^2) \left[
C_0\left(m^2,s,m^2;\underline{m}^2,0,\sigma \right)  
+ 
\frac{B_0\left(m^2;\underline{m}^2,0\right)}{\sigma-s}
\right].
\nl
\ea
The dotted functions have been solved with the aid of {\tt IdSolver}:
\ba
\label{bh-48c}
 B_0\left(m^2;\underline{m}^2,0\right)
&=&
\frac{D-2}{2}\frac{A_0(m^2)}{m^2},
\\
\label{bh-48k}
C_0\left(m^2,s,m^2;\underline{m}^2,0,\sigma \right) &=& 
\epsilon~\frac{s}{m^2(\sigma-s)}~C_0\left(m^2,s,m^2;{m}^2,0,\sigma \right)  
\nl&&
-~ (1-\epsilon)
\Biggl[
\frac{\sigma(s-2m^2)-2m^2s}{m^2(\sigma-4)(\sigma-s)^2}\frac{A_0(\sigma)}{\sigma}
+\frac{\sigma-2m^2}{2m^2(\sigma-4m^2)(\sigma-s)}
\frac{A_0(m^2)}{m^2}
 \Biggr]
\nl&&
+~ (1-2\epsilon)\left[\frac{s}{m^2(\sigma-s)^2}B_0(s;\sigma,0)
-\frac{1}{(\sigma-4m^2)(\sigma-s))}B_0(m^2;\sigma,m^2)
\right].
\ea
The equations (\ref{bh-48c})--(\ref{bh-48k}) are fulfilled for any
$D$, and the 
singularities of (\ref{bh-48d}) in $\epsilon$ come only from the
product of $B_0$ functions.
We show the first terms of the $\epsilon$-expansion for $m^2=1$:
\ba
\label{v4l1}
{\tt V4l3md}
&=& \left[\frac{1}{2\epsilon} + \frac{\zeta_2}{4}\epsilon + \ldots 
\right]
\Biggl[
\frac{1}{\epsilon} + \left(2+\frac{1+x}{1-x}\ln(x) \right)
+
\Biggl\{
2\frac{1+x}{1-x} \left[-{\litwo}(-x) +\ln(x)
  \left(1-\ln(1-x)+\frac{1}{4}\ln(x) \right) \right] 
\nl &&
+ \frac{2 \zeta_2}{1-x} +\frac{1}{2}\left(3\zeta_2+8\right)
\Biggr\}\epsilon 
+ \ldots 
\Biggr]
\nl &&
+~ \int_{4}^{\infty}\frac{d\sigma}{\sigma-s}
\left[\ln\frac{\sqrt{\sigma}-\sqrt{\sigma-4}}{2} 
- \frac{1}{2}\sqrt{\sigma(\sigma-4)}\left[\ln(\sigma) +
  2\ln\left(1-\frac{s}{\sigma}\right) \right] 
+\ldots
\right] + {\cal O}(\epsilon) .
\ea
The result agrees with  {\tt V4l3md} given in the file
{\tt MastersBhabha.m} \cite{web-masters:2004nn}, where it was derived
with a differential equation.
The first two coefficients are easily read from (\ref{v4l1}), and the
constant term is: 
\ba
{\tt V4l3md}[0,x] &=&
2 - \frac{\zeta_2}{1-x} + \frac{1+7x}{4(1-x)} \ln^2(x) 
+ \frac{1+x}{1-x}\left\{
\ln(x)\left[1-\ln(1-x)-3\ln(1+x)\right] -3{\litwo}(-x) - {\litwo}(x)
\right\} .
\nl
\ea

\subsection{\label{subboxes}Two-loop 4-point Master Integrals}
There are thirty three two-loop box MIs.
They are shown in Figures \ref{beta6} to \ref{beta8}.
Several analytical expressions for box MIs in terms of HPLs may be found in
the file {\tt MastersBhabha.m} \cite{web-masters:2004nn}. 
In Table \ref{tab-boxes} we list the double-box
diagrams to which they contribute.
Additionally, one may see the other contributing masters.
By now, few of the two-loop box MIs are known analytically.
For diagram B1, the MIs {\tt B7l4m1} and {\tt B7l4m1N} are  given
in \cite{Smirnov:2001cm}; 
for diagram B2, the MI {\tt B7l4m2} is known as
two-dimensional integral representation \cite{Heinrich:2004iq};
for diagram B3, the leading
divergent part of  MI {\tt B7l4m3} is published \cite{Heinrich:2004iq}.
The technique used 
is the
Mellin-Barnes representation
\cite{Boos:1990rg,Smirnov:2004ip,Smirnov:book4} in combination with
summation techniques {\it \`a la}
\cite{Vermaseren:1998uu,Moch:2001zr,Moch:2002hm}. 
Other box MIs were derived with systems of differential equations.
For diagrams B1 and B3 the MI {\tt B5l2m1} has been given in
\cite{Czakon:2004tg}. 
For diagram B5, a set of two double-box MIs, {\tt B5l4m} and  {\tt
  B5l4mN},  was derived recently with the restriction to electrons and
photons
(one flavor) in \cite{Bonciani:2003cj,web-masters:2004nn}.
For our set of masters, we use instead of  {\tt B5l4mN} a dotted function:
\begin{eqnarray}
{\tt B5l4md}  &=& 
\frac{-x y} {(1 - x^2) (1 - y^2)} 
\Bigl[ \frac{ H[0, x] H[0, y] }{\epsilon}
+ G[-1/y   , 0, 0, x] - G[-y, 0, 0, x] -
  \zeta_2 H[0, x] 
\nl &&
-~ 4 \zeta_2 H[0, y] + G[-1/y   , 0, x] H[0, y] +
  G[-y, 0, x] H[0, y] 
- 2 H[0, y] H[-1, 0, x]
\nl && 
-  6 H[0, x] H[-1, 0, y] +
  4 H[0, x] H[0, 0, y] + G[-1/y   , x] [3 \zeta_2 + H[0, 0, y]] 
\nl &&
- G[-y, x] [3 \zeta_2 + H[0, 0, y]] + 2 H[0, x] H[1, 0, y] - H[0, 0, 0, y] 
\bigr] 
+ {\cal O}(\epsilon).
\label{ex}
\end{eqnarray}
The expression depends on HPLs $H$, but also on two-dimensional HPLs
$G$ (see 
Appendix A of \cite{Gehrmann:2000zt}; we use here the notations of
\cite{Bonciani:2003cj} and the $G[-y,0,0,x]$ as given in
\cite{Czakon:2004tg}). 

A differential operator for the derivation of differential equations
for box diagrams in the $s$-channel is: 
\begin{eqnarray}
s \frac{\partial }{\partial s}  = \frac12 \left\{
\left(p_1^\mu +  p_2^\mu\right) + \frac{s\left( p_2^\mu -p_3^\mu
  \right)}{s+t-4}  
    \right\} \frac{\partial}{\partial p_2^\mu} .
\label{sop}
\end{eqnarray}
The corresponding operator in the $t$-channel is obtained by the replacements
$ s  \leftrightarrow t$ and $p_2 \leftrightarrow -p_3$.
Since the number of differential operators  $p_i^\mu \partial /
\partial p_j^\mu$ is larger than the number of kinematic invariants,
 there is some freedom of choice.
 Let us note that 
representation (\ref{sop}) is much simpler than {\em e.g.} that chosen  
in  \cite{Bonciani:2003cj}
 \footnote{Note the misprint in Equation (14) of \cite{Bonciani:2003cj}. }.

\begin{figure}[ht]
\epsfig{file=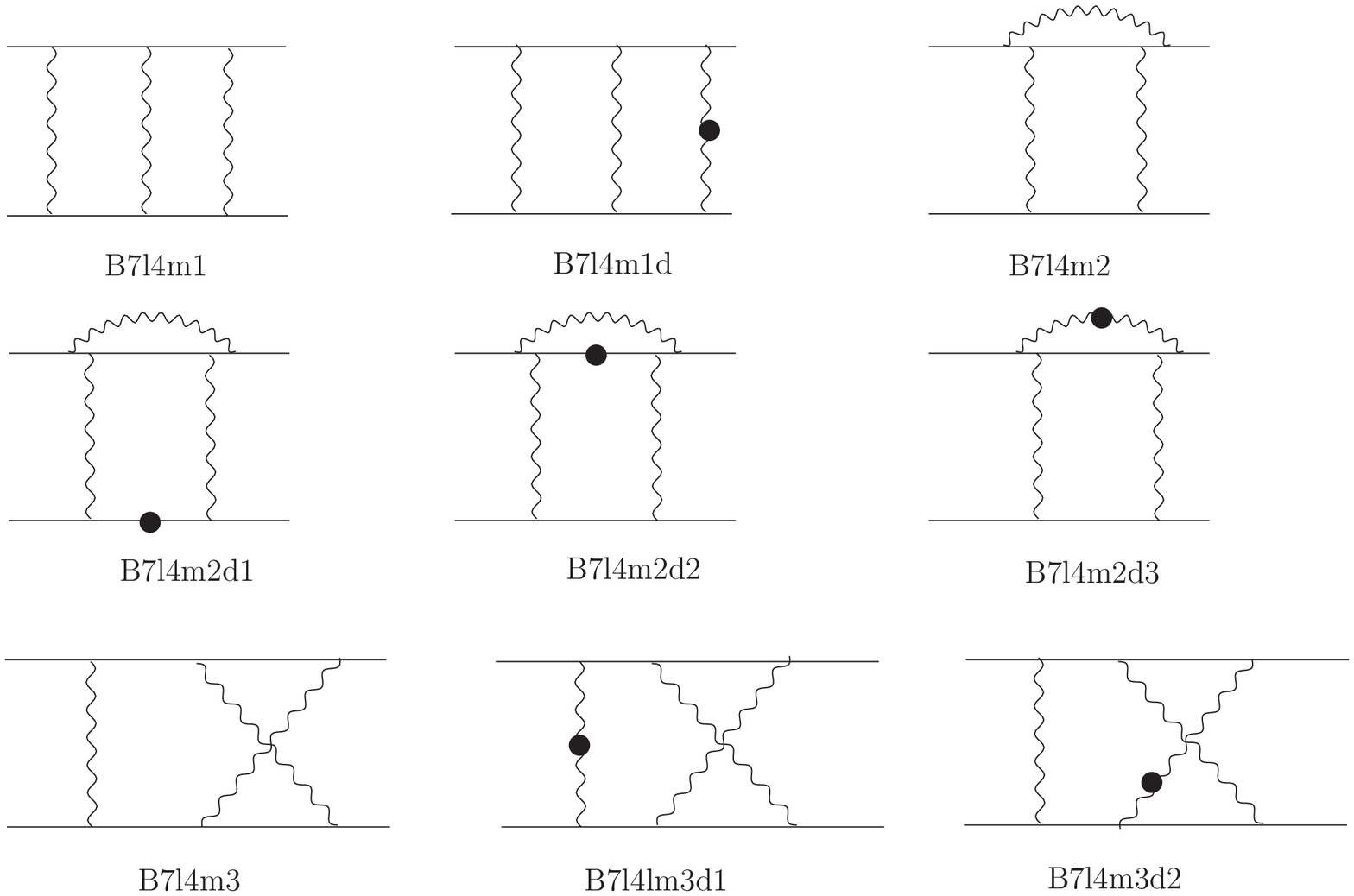, width=12cm}
\caption{The nine  two-loop box MIs with seven internal lines.}
\label{beta6}
\end{figure}

\begin{figure}[ht]
\epsfig{file=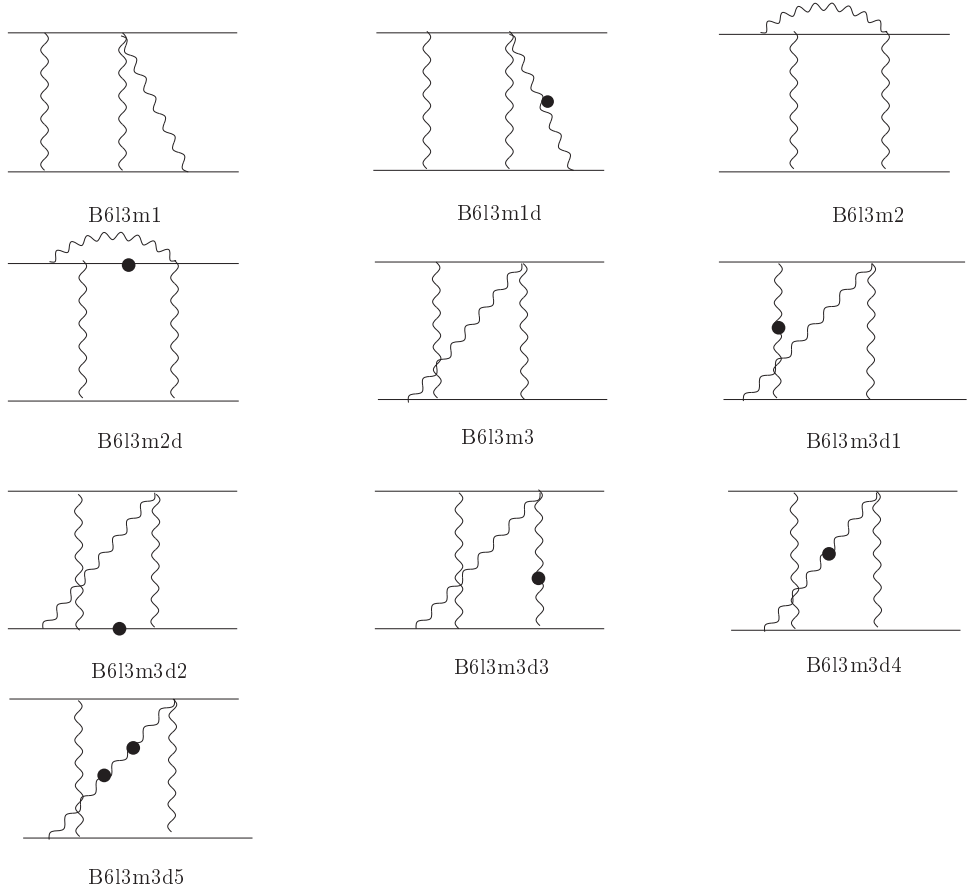, width=13cm}
\caption{The ten  two-loop box MIs with six internal lines.}
\label{beta7}
\end{figure}

\begin{figure}[ht]
 \epsfig{file=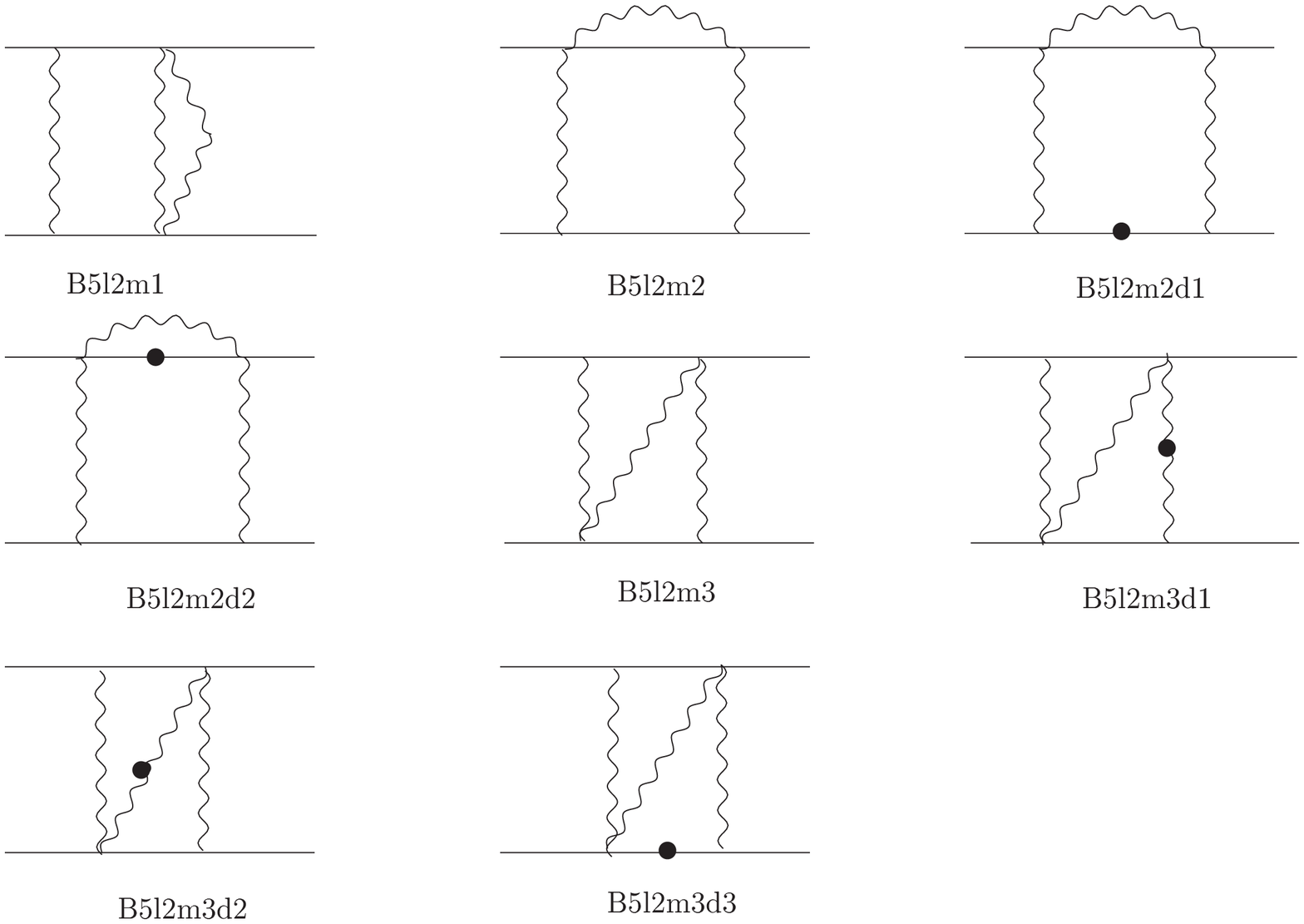, width=13cm}
\\
\epsfig{file=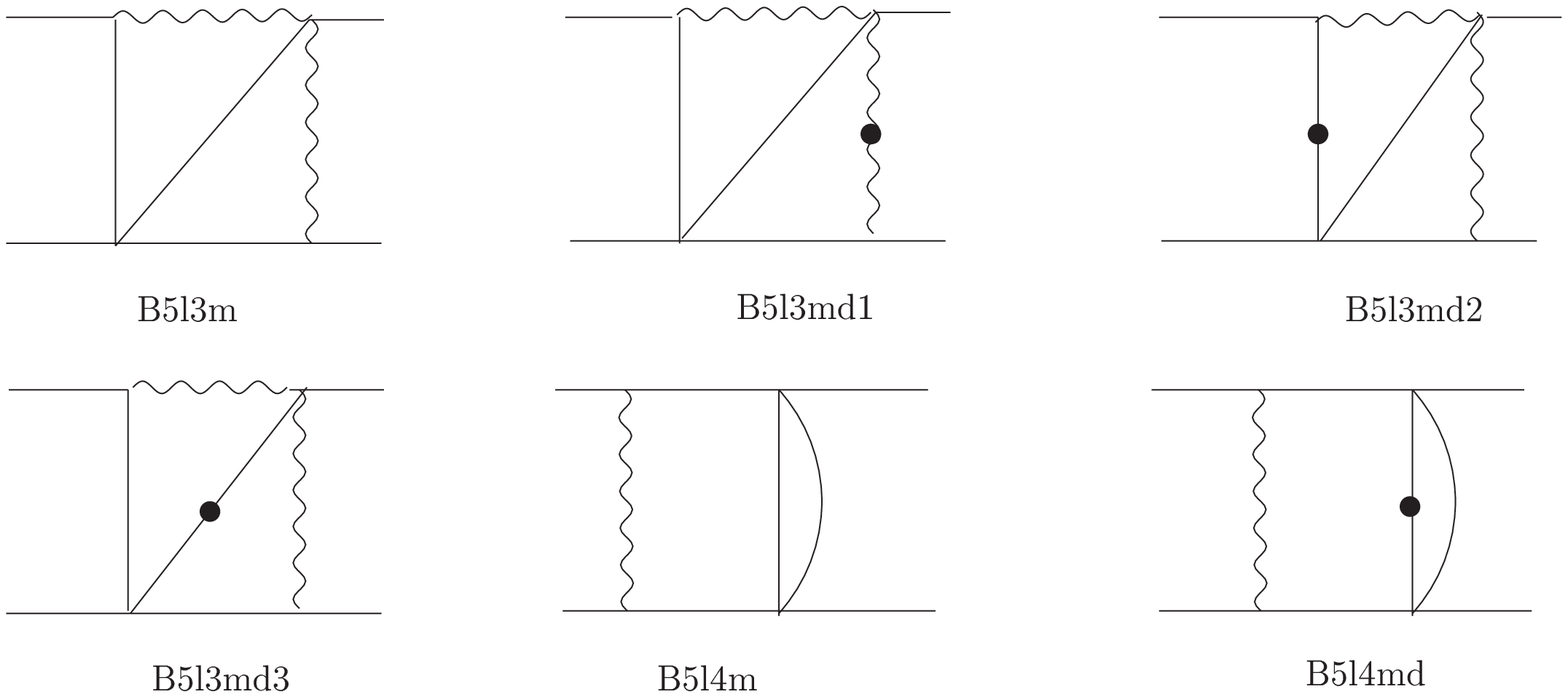, width=13cm}
\caption{The fourteen two-loop box MIs with five internal lines.}
\label{beta8}
\end{figure}

\subsection{\label{sec-numerators}Dotted MIs and MIs with irreducible
  numerators} 
From the point of view of automatic calculations, there 
is no essential difference whether a set of MIs is given with or
without  MIs with irreducible numerators.
There exist algebraic relations to transform from one set to the other.
However, when the MIs are determined 
with the help of differential  equations, the solution may come out
much easier when the unknown MIs are properly chosen. 
Introducing numerators or dots on lines will  change the
nature of a MI concerning both the UV and IR singularities.
Related to this is that some
coupled differential equations for  
MIs can be decoupled due to separated orders of their $\epsilon$-expansions.
Moreover, the solution for a given MI can be much simpler 
compared to that for another choice. 
As an example
may serve (\ref{ex}), which  is by far more compact than the
corresponding MI {\tt   B5l4mN}  with numerator, given in
Equations (37)--(39) of \cite{Bonciani:2003cj}.   
Let us also note that the $\epsilon^2$ singularity of  {\tt   B5l4mN}
is absent in (\ref{ex}).
Contrary,
the masters of  prototype {\tt SE3l2m} \cite{web-masters:2004nn} are
an example 
for the opposite case where a solution with numerator is simpler
than that with a dotted propagator.

Let us finally mention that any of the dotted MIs 
can be replaced by a MI with an appropriate numerator. 
However, it is not always possible to ``move'' a dot arbitrarily from
one line to another line.
{\em E.g.} the dot in {\tt V5l2m2d} cannot be 
moved from the massless line to one of the two internal massive lines.
The arising integral  {\tt V5l2m2} [2, 1, 1, 1, 1;0, 0] (defined in 
(\ref{betav5})) may be considered as a master integral, but 
cannot replace  {\tt V5l2m2d}.
This may be easily proven by  the following two algebraic relations:
\ba
\label{v5l2m2aux1}
 {\tt V5l2m2} [1, 1, 1, 2, 1;0, 0]
&=&
 {\tt V5l2m2d}
- \frac{3(1-5\ep+6\ep^2)}{2x(1+2\ep)} {\tt V4l1m2} 
\nl&&+~
\frac{3(-1+2\ep)(-2+3\ep)(-1+3\ep)}{4x(-1+2\ep+8\ep^2)} {\tt SE3l1m}
+ \frac{3(2-13\ep+27\ep^2-18\ep^3)}{2\ep x^2(1+2\ep)} {\tt SE3l0m},
\\
\label{v5l2m2aux2}
 {\tt V5l2m2} [2, 1, 1, 1, 1;0, 0]
&=&
- \ep ~   {\tt V5l2m2} 
- \frac{1-5\ep+6\ep^2}{2\ep x}  {\tt V4l1m2} +
\frac{(2-13\ep+27\ep^2-18\ep^3)}{4\ep x (1-4\ep)}  {\tt SE3l1m}.
\ea
We conclude for this specific example that the set of masters could be
chosen to contain  
one master out of the pair of integrals with a dot on a massless line
( {\tt V5l2m2d},  {\tt V5l2m2} [1, 1, 1, 2, 1;0, 0]),
and one master out of the pair of integrals with no dot or a dot on a
massive line
( {\tt V5l2m2},  {\tt V5l2m2} [2, 1, 1, 1, 1;0, 0]).
 
\subsubsection{\label{sssecNum}%
The divergent parts of the master integrals for the
  prototypes {\tt B5l2m2} and {\tt B5l2m3}}
As was explained in Section \ref{a1-3}, under certain conditions one
may determine 
singular parts of MIs in a purely algebraic way or with a combination
of algebraic relations and differential equations.
We have used this method in order to determine the singularities
of the so far unknown MI for prototypes {\tt B5l2m2} and  {\tt B5l2m3}:
\begin{eqnarray}
\label{beta6d}
{\tt B5l2m2} &=& \frac{1}{\epsilon} \frac{x}{x^2-1} \left(4 \zeta_2 + H[0, 0,
x] + 2 H[0, 1, x] \right) + {\cal{O}}(1), 
\\
\label{beta6f}
{\tt B5l2m2d1} & = & -\frac{x}{(1 - x)^2} \left[ \frac{1}{\epsilon^2}
  + \frac{1}{\epsilon}  \left( 2 + H[0, x] + 2 H[1, x]\right) \right]
+{\cal{O}}(1) ,
\\
\label{beta6g}
{\tt B5l2m2d2} &=&  {\cal{O}}(1),
\\
\label{beta6h}
{\tt B5l2m3} &=&   {\cal{O}}(1),
\\
\label{beta6j}
{\tt B5l2m3d1} &=& -\frac{1}{\epsilon^2} \frac{ (1 + x^2) y}{8 x (1 -
  y)^2}  
\\
&&-~ \frac{1}{\epsilon} \frac{y}{4 x (1 - y)^2 (1 + y)^3} 
\bigl[ -(1 + x^2) (1 + y)^3  +\zeta_2 \left[ 8 (1 - x)^2 (1 - y) y
\right] 
\\
&&+~ 2 (1 + y) 
\left[ x (1 - y)^2 + 2 y\left(1 + x^2\right) \right] 
\left(H[0,y] +2 H[1,y] \right)
\\
&&-~   (1 - x^2) (1 + y)^3 H[0,x]
+2 y (1 - x)^2 (1 - y)  \left(H[0,0,y] +2 H[0,1,y]\right)
\bigr] +  {\cal{O}}(1),
\\  
\label{beta6k}
{\tt B5l2m3d2} 
&=& 
-\frac{1}{\epsilon^2} \frac{x y H[0, x]}{(1 - x^2) (1 -
  y)^2}
\nl&&
-~
\frac{1}{\epsilon} \frac{2x y}{(1- x^2) (1 - y)^2}
\Bigl(  
H[1,0,x] -H[-1,0,x] + H[0,0,x] 
\nl
&&+~\frac{\zeta_2}{2} + 
H[0,x]\left( H[0,y]+2~H[1,y]\right)
\Bigr)  + {\cal{O}}(1),
\\ 
\label{beta6l}
{\tt  B5l2m3d3} &=& -\frac{1}{\epsilon}  \frac{y}{2 (1 - y^2)} \left[ 4  
  \zeta_2 +    H[0, 0, y] +  2   H[0, 1, y] \right]   + {\cal{O}}(1).
\end{eqnarray}

\subsection{\label{sse-4}Additional master integrals with several flavors }
In QED,
there are additional fermion flavors besides electrons.
In two-loop Bhabha scattering this leads to the additional diagrams
SE5f, V4f, B5f shown in Figure \ref{betaN}.
They may be derived from Figures \ref{SEprot} to \ref{LL2} by a
replacement of the closed 
electron loop in diagrams SE5, V4, B5 by a loop with a second  mass scale
$M$. 
As a consequence, additional MIs will arise.
We have determined them with {\tt DiaGen/IdSolver}, and Table \ref{2loopmastnf2}
lists the MIs 
which contribute to the evaluation of Feynman integrals of the new
prototypes.       

\begin{figure}[htbp]
\epsfig{file=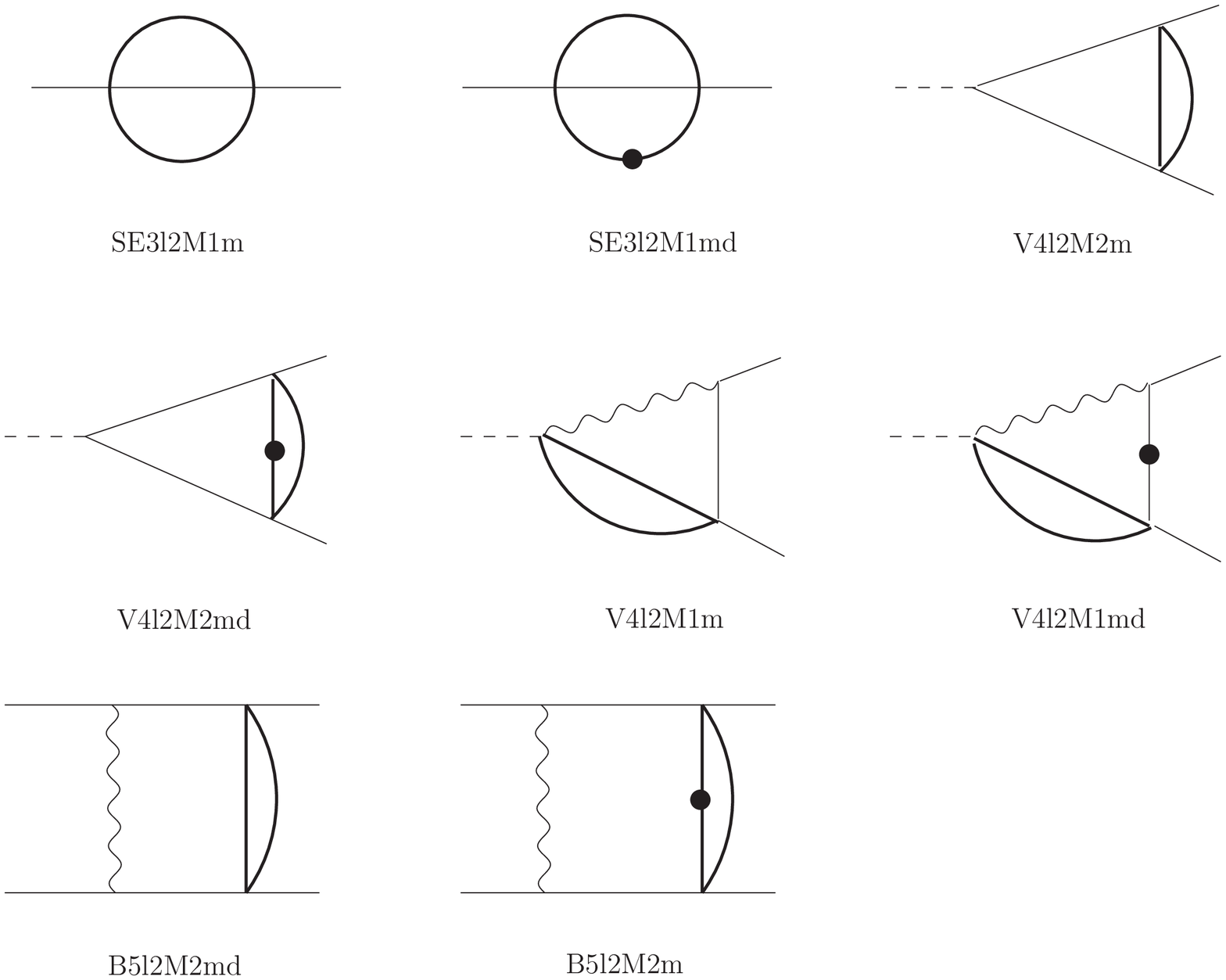, width=12cm}
\caption{The eight additional master integrals with two
  different mass   scales.
}
\label{betaN}
\end{figure}

\section{Summary}
The results presented here 
document a further step towards
a complete two-loop prediction of small angle Bhabha
scattering, needed {\em e.g.} for a precise luminosity determination
at the future International Linear Collider.

We have determined a complete set of scalar master integrals needed
for the calculation of the virtual two-loop corrections to massive Bhabha
scattering in QED, both for one flavor and for several flavors.
The MIs are shown pictorially, and we tabulate to which prototype
diagrams they will contribute. 
Further, we determined the first terms of the $\epsilon$-expansions for
all the two-loop self-energy and vertex master integrals for the
one-flavor case, and some of the two-loop box masters.
Some of these master integrals were unknown so far. 
The analytical expressions are collected in the Mathematica file {\tt
MastersBhabha.m} and are publicly available \cite{web-masters:2004nn}.
As by-products of our numerous tests of the results presented here, we
expressed HPLs up to weight 4 by a 
minimal basis (Mathematica file {\tt HPL4.m}
\cite{web-masters:2004nn}), and have generalized the sector 
decomposition algorithm for the evaluation of Feynman integrals in the
Euclidean region to the case of irreducible numerators.  

Once the last, most complicated master integrals are determined one
will have to combine the virtual two-loop corrections to Bhabha
scattering, together with their
counter terms, with electroweak corrections and with a package for the
treatment of real bremsstrahlung.

\section*{Acknowledgments}
We would like to thank J. Bl{\"u}mlein, M. Kalmykov and S. Moch for
discussions. 
The work  was supported in part 
  by European's 5-th Framework under contract No. HPRN--CT--2000--00149 (Physics at
  Colliders), 
  by TMR under EC-contract No. HPRN-CT-2002-00311 (EURIDICE), by the
  Sofja Kovalevskaja Award of the Alexander von Humboldt Foundation
  sponsored by the German Federal Ministry of Education and Research,
  by Deutsche Forschungsgemeinschaft under contract SFB/TR 9--03, 
  and by the Polish State Committee for Scientific Research (KBN)
  for the research project in years 2004-2005.

\vfill

\begin{table}[h]
\setlength{\tabcolsep}{0.3pc}
\caption{The number of (two-loop + one-loop) master integrals needed to
  calculate the 
  two-loop vertex diagrams and box diagrams
  (Fig. \ref{SEprot})}
\label{table}
\begin{tabular}{|r|rrrrr|rrrrrr|}
\hline
Diagram      &V1 &V2 &V3  &V4 &V5 &B1  &B2 &B3 &B4  &B5 &B6  \\   
\hline
tadpole MI   &0+1&0+1 &0+1&0+1&0+1&0+1 &0+1&0+1&0+1 &0+1&0+1 \\
2-point MIs  &3+1&4+1 &4+1&1+1&3+1&4+1 &5+2&5+1&4+2 &3+2&3+2  \\
3-point MIs  &4+0&10+0&5+0&2+0&1+0 &7+0  &11+1&13+0 &10+1 &4+1  &4+1  \\
Box type MIs &  &    &   &   &   &9+0&15+0&22+0&11+1&2+1&3+1  \\
\hline
total        &7+2&14+2&9+2&3+2&4+2 &20+2&31+4&40+2&25+5&9+5&10+5  \\
\hline
net          &9  &16  &11  &5  &6   &22  &35  &42  &30  &14  &15    \\
\hline
\end{tabular}
\end{table}

\vfill

\begin{table}[hbtp]
\caption{The one-loop and two-loop MIs needed for the evaluation of the
  self-energy diagrams}
\label{2loopmast3N1}
\begin{tabular}{|l|c|c|c|c|r|c|c|c|c|}
\hline
MI                    &    SE1 &SE2& SE3 & SE4 & SE5 & ref.\\
\hline \hline
{\tt T1l1m}$^\ast$    &   +    &+&  +  &  +  &  +  &
\cite{'tHooft:1972fi}
 \\
\hline
{\tt SE2l2m}$^\ast$   &+       &+&  --  &  --  &  --  & \cite{'tHooft:1972fi}
\\
\hline
{\tt SE3l1m}          &   --   &--&  +  & -- &  --  & oms:{\cite{Fleischer:1999tu}}
\\ 
{\tt SE3l2m}          &     +  &+&  --  &  --  &  --  &
\cite{Fleischer:1999hp,Bonciani:2003te}
\\ 
{\tt SE3l2md}         &     +  &+&  --  &  --  &  --  &
\cite{Fleischer:1999hp},{Sec. \ref{sec-2l2p}}
\\ 
{\tt SE3l3m}          &    --  &--&  --  &  +  &  +  &
oms:{\cite{Broadhurst:1990ei,Broadhurst:1991fi,Fleischer:1999tu}}
\\  
\hline
\end{tabular}
\end{table}

\vfill

\begin{table}[ht]
\label{tab-vertbox}
\caption{The 1-point, 2-point and 3-point MIs entering basic two-loop vertex
and box diagrams in Figures \ref{SEprot} to \ref{LL2}.
Stars denote one-loop MIs. 
}
\begin{tabular}{|l|c|c|c|c|c|c|c|c|c|c|c|r|}
\hline
MI & B1 & B2 & B3 & B4 & B5 & B6 & V1 & V2 & V3 & V4 & V5& ref.\\
\hline \hline
{\tt T1l1m}$^\ast$ &       + &  + &  + &  + &  + &  +  & + &  +  & +  & +  &  + & \cite{'tHooft:1972fi}\\
\hline
{\tt SE2l0m}$^\ast$       & -- &  + &  -- &  + &  + &  +  & -- &  --  & --& --  &  -- & \cite{'tHooft:1972fi}\\
{\tt SE2l2m}$^\ast$       & + &  + &  + &  + &  + &  +  & + &  +  & +&  +  &  + &\cite{'tHooft:1972fi}\\
{\tt SE3l0m} &            + &  -- &  + &  -- &  -- &  --  & -- &  --  & -- &  --  &  -- &{\cite{'tHooft:1972fi,Surguladze:1989ez,Bonciani:2003hc}}\\
{\tt SE3l1m} &            + &  + &  + &  + &  -- &  +  & + &  +  & + &  --&  +  &oms:{\cite{Fleischer:1999tu}}\\
{\tt SE3l2m} &            + &  + &  + &  + &  + &  +  & + &  + &  + &  --  &  + &\cite{Fleischer:1999hp,Bonciani:2003te}\\
{\tt SE3l2md} &           + &  + &  + &  + &  + &  +  & + &  + &  + &  --  &  + &\cite{Fleischer:1999hp},Sec. \ref{sec-2l2p}\\
{\tt SE3l3m} &            -- &  + &  + &  + &  + &  --  & -- &  + &  +
&  +  &  -- &oms: {\cite{Broadhurst:1990ei,Broadhurst:1991fi,Fleischer:1999tu}} \\
{\tt SE5l3m} &            -- &  + &  -- &  -- &  -- &  --  & -- &  -- &  --
&  --  &  -- &{\cite{Broadhurst:1990ei,Fleischer:1998nb,Davydychev:2003mv,Bonciani:2003hc}}\\ 
\hline
{\tt V3l1m}$^\ast$        & -- &  + &  -- &  + &  + &  +  & -- &  -- &-- &  --  &  -- & \cite{'tHooft:1972fi}\\
\hline
{\tt V4l1m1} &            -- &  + &  -- &  + &  -- &  +  & -- &  -- &  -- &  --  &  -- & \cite{Bonciani:2003hc}\\
{\tt V4l1m1[d1--d2]} &     -- &  + &  -- &  + &  -- &  +  & -- &  -- &  -- &  --  &  -- & \cite{Bonciani:2003hc}\\
{\tt V4l1m2} &            + &  -- &  + &  -- &  -- &  --  & -- &  -- &  -- &  --  &  -- & \cite{Bonciani:2003hc}\\
{\tt V4l2m1} &            + &  -- &  + &  -- &  -- &  --  & + &  + &  -- &  --  &  -- & \cite{Bonciani:2003hc}\\
{\tt V4l2m2} &            + &  + &  + &  + &  -- &  +  & + &  + &  + &  -- &  + & \cite{Bonciani:2003te}\\
{\tt V4l3m} &             -- &  + &  + &  + &  + &  --  & -- &  + &  + &  --  &  -- & \cite{Bonciani:2003te}\\
{\tt V4l3md} &            -- &  + &  + &  + &  + &  --  & -- &  + &  + &  --  &  -- & \cite{Bonciani:2003te}\\
{\tt V4l4m} &             -- &  + &  + &  + &  + &  --  & -- &  + &  + &  +  &  -- & \cite{Bonciani:2003te}\\
{\tt V4l4md} &            -- &  + &  + &  + &  + &  --  & -- &  + &  + &  +  &  -- & \cite{Bonciani:2003te}\\
\hline
{\tt V5l2m1} &            -- &  + &  -- &  + &  -- &  --  & -- &  -- &  -- &  --  &  -- & \cite{Czakon:2004tg}\\
{\tt V5l2m1d} &           -- &  + &  -- &  + &  -- &  --  & -- &  -- &  -- &  --  &  -- & \cite{Czakon:2004tg}\\
{\tt V5l2m2} &            + &  -- &  + &  -- &  -- &  --  & -- &  -- &  -- &  --  &  -- & \cite{Czakon:2004tg}\\
{\tt V5l2m2d} &           + &  -- &  + &  -- &  -- &  --  & -- &  -- &  -- &  --  &  -- & \cite{Czakon:2004tg}\\
{\tt V5l3m} &             + &  -- &  + &  -- &  -- &  --  & + &  + & -- &  --  &  -- & \cite{Bonciani:2003te}\\
{\tt V5l3md} &            + &  -- &  + &  -- &  -- &  --  & + &  + & -- &  --  &  -- & \cite{Bonciani:2003te}\\
\hline
{\tt V6l4m1} &            -- &  -- &  + &  -- &  -- &  --  & -- &  + &  -- &  --  &  -- & \cite{Bonciani:2003te}\\
{\tt V6l4m1d} &           -- &  -- &  + &  -- &  -- &  --  & -- &  + &  -- &  --  &  -- & \cite{Bonciani:2003te}\\
{\tt V6l4m2} &            -- &  + &  -- &  -- &  -- &  --  & -- &  -- &  -- &  --  &  -- & \cite{Bonciani:2003hc}\\
\hline
total = 25+4$^\ast$ &  11+2$^\ast$ & 16+4$^\ast$ & 16+4$^\ast$ &  14+4$^\ast$& 7+4$^\ast$ & 7+4$^\ast$  & 7+2$^\ast$ &  14+2$^\ast$& 9+2$^\ast$ &  3+2$^\ast$   & 4+2$^\ast$ & \\
\hline
\end{tabular}
\end{table}

\vfill

\begin{table}[htbp]
\label{tab-boxes}
\caption{
4-point MIs entering basic two-loop box diagrams in Figure \ref{LL2}. 
An asterisk  denotes one-loop MI.
MIs with a dagger are not known completely analytically; see text.
}
\begin{tabular}{|l|c|c|c|c|c|c|l|}
\hline
MI &                B1 & B2 & B3 & B4 & B5 & B6& ref.\\
\hline \hline
{\tt B7l4m1} &            +  &  -- & --  & --  &  -- & -- & \cite{Smirnov:2001cm} \\
{\tt B7l4m1N} &           +  &  -- & --  & --  &  -- & --
&\cite{Heinrich:2004iq}
\\ 
{\tt B7l4m2} &            --
&  + & --  & --  &  -- & --  & 
\cite{Heinrich:2004iq}$^{\dagger}$
\\ 
{\tt B7l4m2[d1--d3]} &     --  & +  & --  & --  &  -- & --
&\\ 
{\tt B7l4m3}  &            --  & --  &
+  & --  &  -- & --  & 
\cite{Heinrich:2004iq}$^{\dagger}$
\\
{\tt B7l4m3[d1--d2]} &     --  & --  & +  & --  &  -- & --  &\\

\hline
{\tt B6l3m1} &            +  & --  & +  & --  &  -- & --  &\\
{\tt B6l3m1d} &           +  & --  & +  & --  &  -- & --  &\\
{\tt B6l3m2} &            --  & +  & --  & +   &  -- & --  &\\
{\tt B6l3m2d} &           --  & +  & --  & +   &  -- & --  &\\
{\tt B6l3m3} &            --  & --  & +  & --  &  -- & --  &\\
{\tt B6l3m3[d1--d5]} &     --  & --  & +  & --  &  -- & --  &\\
\hline
{\tt B5l2m1} &            +  & --  & +  & --  &  -- & -- & \cite{Czakon:2004tg} \\
{\tt B5l2m2}  &           --  & +  & --  & +  &  -- & +
&{Sec. \ref{sssecNum}}$^{\dagger}$\\ 
{\tt B5l2m2[d1--d2]} &     --  & +  & --  & +  &  -- & +
&{Sec. \ref{sssecNum}}$^{\dagger}$\\ 
{\tt B5l2m3}    &         +  & --  & +  & --  &  -- & --  &\\
{\tt B5l2m3[d1--d3]} &     +  & --  & +  & --  &  -- &
--&{Sec. \ref{sssecNum}}$^{\dagger}$\\ 
{\tt B5l3m}     &        --  & +  &  +  & +  & --  & --  &\\
{\tt B5l3m[d1--d3]}&      -- & +  &  +  & +  & --  & --  &\\
{\tt B5l4m}    &          --  & +  & +  & +  &  + & --  & \cite{Bonciani:2003cj} \\
{\tt B5l4md} &            --  & +  & +  & +  &  + & --  & Sec. \ref{subboxes}
\\
\hline
{\tt B4l2m}$^\ast$&       --  & --  & --  & +  &  + & +  & \cite{'tHooft:1972fi,Bonciani:2003cj}  \\
\hline
total = 33+1$^\ast$ &       9  & 15  & 22  & 11+1$^\ast$  &  2+1$^\ast$ & 3+1$^\ast$ & \\
\hline
\end{tabular}
\end{table}

\vfill

\begin{table}[hbtp]
\caption{The two-loop master integrals for several flavors. An asterisk
  denotes a one-loop MI.}
\label{2loopmastnf2}
\begin{tabular}{|l|c|c|c|r|c|c|}
\hline
MI           & SE5f &  V4f  & B5f  & ref.\\
\hline \hline
{\tt T1l1m}$^\ast$     & +   &  +  &  +  &  \cite{'tHooft:1972fi}\\
\hline
{\tt SE2l2m}$^\ast$    & --  &  +  &  + & \cite{'tHooft:1972fi}\\
{\tt SE2l0m}$^\ast$    & --  &  --  &  + & \cite{'tHooft:1972fi}\\
\hline
{\tt SE3l1m}           & --  &  --  &  +  & oms:{\cite{Fleischer:1999tu}} \\
{\tt SE3l2m}           & --  &  --  &  +  &\cite{Fleischer:1999hp,Bonciani:2003te}\\
{\tt SE3l2md}          & --  &  --  &  +  &\cite{Fleischer:1999hp},{Sec. \ref{sec-2l2p}}\\
{\tt SE3l2M1m}          &+ &  + &  +  & oms:\cite{Argeri:2002wz} \\
{\tt SE3l2M1md}& +  &  + &  +  & oms:\cite{Argeri:2002wz} \\
\hline
{\tt V3l1m}$^\ast$             & --   &  --  &  + & \cite{'tHooft:1972fi}\\
{\tt V4l2M1m}           & --   &  --  &  + & \\
{\tt V4l2M1md}          & --   &  --  &  + & \\
{\tt V4l2M2m}           & --   &   +  &   + & \\
{\tt V4l2M2md}          & --   &  +  &   + & \\
\hline
{\tt B4l2m}$^\ast$             & --   &  --  &  + &\cite{'tHooft:1972fi,Bonciani:2003cj} \\
{\tt B5l2M2m}           & --  &  --  &  + & \\
{\tt B5l2M2md}          & --   &  --  &  + & \\
\hline
\end{tabular}
\end{table}

\clearpage

\appendix

\section{\label{sec-a2}Numerical evaluation of
  multi-loop diagrams with irreducible numerators 
in the Euclidean region} 
Here we derive expressions  for the Feynman integrals $G(X)$ defined in
 (\ref{bh-48}) with $L$
loops, a numerator function $X(k_i)$, N internal lines, and $E$
external legs with momenta $p_e$ (see also {\em e.g.}
\cite{Nakanishi:1971}).  
The propagator momenta $q_i$ compose as: 
\ba 
\label{bh-7}
q_i &=& c_1^ik_1 + \ldots c_L^ik_L + \ldots + d_1p_1 + \ldots + d_E^ip_E.
\ea
In 
\cite{Binoth:2000ps,Binoth:2003ak},
a method was derived for the numerical evaluation of multi-loop
Feynman integrals in the Euclidean region.
The method is based on sector decomposition (see also
\cite{Hepp:1966eg,Roth:1996pd}). 
In \cite{Binoth:2000ps,Binoth:2003ak}, the method was worked out explicitely
for propagators only, 
including dotted ones ($\nu_i \neq 1$), {\em i.e.} for $X=1$.
We need the formulae also with numerators and repeat the starting
expression for completeness here, in our notations: 
\bea
  \label{bh-19a}
G(X)
&=& 
\frac{1}{(i\pi^{d/2})^L} \int \frac{d^Dk_1 \ldots d^Dk_L~~X}
     {(q_1^2-m_1^2)^{\nu_1} \ldots (q_j^2-m_j^2)^{\nu_j} \ldots
       (q_N^2-m_N^2)^{\nu_N}  }  .
\ea
The denominator of $G(1)$ contains, after introduction of Feynman
parameters $x_i$, the momentum dependent function
\ba
\label{bh-6}
m^2 &=& \sum_{i=1}^{N} x_i (q_i^2-m_i^2) ~=~ kMk - 2kQ + J.
\ea
The linear terms in $m^2$ may be eliminated by a shift:
\ba
\label{bh-8}
k &=& {\bar k} + M^{-1} Q,
\ea
and the matrix $M$ in (\ref{bh-6}) may be diagonalized by a rotation:
\ba
\label{bh-30}
k \to k' &=& V~k,
\\
\label{bh-31}
kMk &=& k'M_{diag}k'
\\
&\to& \sum \alpha_i k_i^2,
\\
\label{bh-32}
M_{diag} &=& (V^{-1})^+MV^{-1} ~~=~ (\alpha_1,\ldots,\alpha_L) .
\ea
After the diagonalization, the momentum integrals may be easily
performed, and the remaining Feynman parameter
integral contains the term:
\bea
  \label{bh-22}
\frac{(\text{det}~ M)^{-D/2}}
{\left(\mu^2\right)
^{N_{\nu}-DL/2}}
&=& \frac{U(x)^{N_{\nu}-D(L+1)/2}}{F(x)^{N_{\nu}-DL/2}},
\eea
with
\bea
\label{bh-11}
\mu^2 &=& - (J-QM^{-1}Q),
\\
\label{bh-3}
N_{\nu} &=& \nu_1 + \ldots \nu_N ,
\eea
and the definitions \footnote{We deviate from
  \cite{Binoth:2000ps} by a sign in the   definition of $F(X)$.}: 
\ba
U{(x)} &=& (\text{det}~ M),
\\
F({x}) &=& (\text{det}~ M) \mu^2 ~=~ -(\text{det}~ M)~J + Q\tilde{M}Q.
\ea
The $\tilde{M}$ is defined with $M^{-1}=\tilde{M}/\text{det}~ M$.
Finally, one arrives at:
\bea
\label{bh-19}
G(1)&=& (-1)^{N_{\nu}}
\frac{\Gamma\left(N_{\nu}-\frac{D}{2}L\right)}
{\Gamma(\nu_1)\ldots\Gamma(\nu_N)} 
\int_0^1 \prod_{j=1}^N dx_j ~ x_j^{\nu_j-1}
\delta\left(1-\sum_{i=1}^N x_i\right) 
\frac{U(x)^{N_{\nu}-D(L+1)/2}}{F(x)^{N_{\nu}-DL/2}}.
\eea
Equation (3) in \cite{Binoth:2000ps} contains formula (\ref{bh-19}) for
$\nu_i=1$.

From the above it is straightforward to formulate the parameter
Feynman parameter integrals for the case of nontrivial numerators.
The corresponding formulae for the simplest cases are: 
\bea
\label{bh-24a}
G(k_{1\alpha}) &=& (-1)^{N_{\nu}}
\frac{\Gamma\left(N_{\nu}-\frac{D}{2}L\right)}
{\Gamma(\nu_1)\ldots\Gamma(\nu_N)} 
\int_0^1 \prod_{j=1}^N dx_j ~ x_j^{\nu_j-1}
\delta\left(1-\sum_{i=1}^N x_i\right) 
\frac{{U(x)^{N_{\nu}-1-D(L+1)/2}}}{{F(x)}^{N_{\nu}-DL/2}}
\left[\sum_{l} {\tilde{M}}_{1l}Q_l\right]_{\alpha} ,
\nonumber \\
\eea
and
\bea
 \label{bh-47}
G(k_{1\alpha} k_{2\beta})
&=&
(-1)^{N_{\nu}}  
\frac{\Gamma\left(N_{\nu}-\frac{D}{2}L\right)}
{\Gamma(\nu_1)\ldots\Gamma(\nu_N)} 
\int_0^1 \prod_{j=1}^N dx_j ~ x_j^{\nu_j-1}
\delta\left(1-\sum_{i=1}^N x_i\right) 
\frac{{U(x)^{N_{\nu}-2-D(L+1)/2}}}
{{F(x)}^{N_{\nu}-DL/2}}
\nl
&&\times
\sum_{l}\Biggl[
 [{\tilde{M}}_{1l}Q_l]_{\alpha}[{\tilde{M}}_{2l}Q_l]_{\beta}
-
\frac{\Gamma\left(N_{\nu}-\frac{D}{2}L-1\right)}
{\Gamma\left(N_{\nu}-\frac{D}{2}L\right)}
\frac{g_{\alpha\beta}}{2} {U(x)F(x)}
 \frac{(V_{1l}^{-1})^+ (V_{2l}^{-1})}   {\alpha_l}
\Biggr].
\nl
\eea
The general case is treated in \cite{Denner:2004iz}.
Potential UV singularities arise from the overall factors with
$\Gamma$ functions and from $U(x)$, while IR singularities
arise from $F(x)$.
The Mathematica code {\tt sectors.m} \cite{Czakon:2004ww} performs the
isolation of IR singularities by sector decomposition and the
integrations in the Euclidean region. 
For the numerical calculations the package {\tt CUBA}
\cite{Hahn:2004fe} is used. 

\section{\label{a-hpl}Harmonic polylogarithms up to weight 4}
The master integrals in the file
{\tt MastersBhabha.m} \cite{web-masters:2004nn}
are expressed by harmonic polylogarithms (HPLs) \cite{Remiddi:1999ew} up to
weight 4.  
Harmonic polylogarithms fulfill simplified algebraic relations of harmonic
sums.
One may use this and determine a basis of functions
by direct evaluation, and then express the others by algebraic relations.
For details we refer to \cite{Blumlein:2003gb} and references therein.   
 
The three HPLs of weight 1 are:
\ba
\label{hp1-1}
H[0,x] &=& \ln(x),
\\
\label{hp1-2}
H[1,x] &=& -\ln(1-x),
\\
\label{hp1-3}
H[-1,x] &=& \ln(1+x).
\ea

The nine HPLs of weight 2 may be expressed by three independent HPLs of
weight 2 plus those of lower weight: 
\ba
\label{hp2-2}
H[0,1,x] &=& {\litwo}(x),
\\
\label{hp2-3}
H[0,-1,x] &=& -{\litwo}(-x),
\\
\label{hp2-7}
H[-1,1,x] &=& {\litwo}\left(\frac{1+x}{2}\right)
-{\litwo}\left(\frac{1}{2}\right) -\ln2~H[-1,x]. 
\ea
The other HPLs  are determined with the relation
$H[a,b,x] = - H[b,a,x] + H[a,x] H[b,x]$. 

There are 27 HPLs of weight 3 \cite{Moch:1999eb}, and 8 of them have to
be added to the basis:
one  for each of the six index sets of the class $(a,a,b)$ (see
equations (3.22), (3.23) of \cite{Blumlein:2003gb}):
\ba
\label{hp3-2}
H[0,0,1,x] &=& {\litr}(x),
\\
\label{hp3-3}
H[0,0,-1,x] &=& -{\litr}(-x),
\\
\label{hp3-25}
 H[ 0, 1, 1,x] &=&  S_{1,2}(x),
\\
\label{hp3-26}
 H[ 0,-1,-1,x] &=&   S_{1,2}(-x),
\\
\label{hp3-288}
 H[-1, 1, 1,x] &=& S_{1,2}\left(\frac{1+x}{2}\right)-
 S_{1,2}\left(\frac{1}{2}\right) 
-
\ln2\left[{\litwo}\left(\frac{1+x}{2}\right)-{\litwo}\left(\frac{1}{2}\right)\right]
 +\frac{1}{2}\ln^2 2~H[-1,x], 
\\
\label{hp3-299}
H[ 1,-1,-1,x] &=& - S_{1,2}\left(\frac{1-x}{2}\right)
+ S_{1,2}\left(\frac{1}{2}\right)  
+ \ln2\left[{\litwo}\left(\frac{1-x}{2}\right)
  -{\litwo}\left(\frac{1}{2}\right)\right] 
 + \frac{1}{2}\ln^2 2 ~ H[1,x], 
\ea
and two for the class $(a,b,c)$  (see
equations (3.7) to (3.10) of \cite{Blumlein:2003gb}):
\ba
\label{hp3-211}
H[ 1, 0,-1,x] &=&  \frac{1}{2} S_{1,2}(x^2)
 - H[0,1,1,x] - H[0,-1,-1,x]
+H[1,x]H[0,-1,x], 
\\
\label{hp3-int}
 H[ 0,-1, 1,x] &=&
 \int_0^x\frac{dy}{y}\left[{\litwo}\left(\frac{1+y}{2}\right)-
 {\litwo}\left(\frac{1}{2}\right)\right]  
- \ln2~H[0,-1,x]. 
\ea
Further, $H[a,a,a,x] = \frac{1}{6} H[a,x]^3$.  
The integral in (\ref{hp3-int}) may be performed analytically, see
$V_3(x)$ in (\ref{hp3-212}) or equation (188) in \cite{Moch:1999eb}. 

There are 81 HPLs of weight 4:
the three index sets of class
$(aaaa)$ have 1 element each (none is basic), the six index sets of
class $(aaab)$ have 4 elements 
each (one is basic), the  three index sets of class $(aabb)$ have 6 elements
each (one is basic), and the three index sets of class $(aabc)$ have
12 elements each (three are basic).
The 18 independent HPLs of weight 4 may be chosen to be:
\ba
\label{hp4-2a}
 H[ 0, 0, 0, 1,x] &=&  {\lifo}( x),
\\
\label{hp4-1g}
 H[ 0, 0, 0,-1,x] &=&  - {\lifo}( -x),
\\
\label{hp4-2d}
H[ 0, 1, 1, 1,x] &=&  S_{1,3}(x),
\\
\label{hp4-2d2}
H[ -1, 1, 1, 1,x] &=& 
-{\lifo} \left(\frac{1-x}{2} \right) +{\lifo} \left(\frac{1}{2} \right) 
+\ln(1-x){\litr}\left(\frac{1-x}{2} \right)
-\frac{1}{2}{\ln}^2(1-x){\litwo}\left(\frac{1-x}{2} \right)
\nl &&
-~\frac{1}{6}\ln^3(1-x)\ln\frac{1+x}{2}
,
\\
\label{hp4-2j}
H[ 0,-1,-1,-1,x] &=& - S_{1,3}(-x), 
\\
\label{hp4-2j2}
H[ 1,-1,-1,-1,x] &=& H[-1,1,1,1,-x], 
\\
\label{hp4-2s}
 H[ 0, 0, 1, 1,x] &=&  S_{2,2}(x),
\\
\label{hp4-2h}
H[ 0, 0,-1,-1,x] &=&  S_{2,2}(-x),
\\
\label{hp4-2h2}
H[ 1, 1,-1,-1,x] &=& \int_0^x \frac{dy}{1-y}H[1,-1,-1,y]
\nl 
&=&
-K_1(x)  
+\ln 2 ~J_2(x) 
+
\frac{1}{2} \ln^2 2~ H[1,1,x],
\\
\label{hp4-1}
H[-1,0,0,1,x] &=&  \int_0^x \frac{dy}{1+y} {\litr}(y)
,
\\
\label{hp4-2}
H[-1,0,1,0,x]  
&=&
-~ \frac{1}{4} S_{2,2}\left(x^2\right)
- H[-1, 0, 0, 1, x] 
+ H[0, 0, -1, -1, x] 
+ H[0, 0, 1, 1, x] 
\nl&&
-~H[-1, x]H[0, 0, 1, x] 
\nl&&
+~ H[0, x]( -H[1, x] H[0, -1, x] + H[-1, x] H[0, 1, x] + H[1, 0, -1, x]) ,
\\
\label{hp4-3}
H[-1,1,0,0,x]&=& - \int_0^x \frac{dy}{1+y} \left[ S_{1,2}(1-y) -
  S_{1,2}(1)  \right]
\nl &=&
-\ln(1+x)   \left[ S_{1,2}(1-x) - S_{1,2}(1)  \right]             
- \frac{1}{2} 
\int_0^x \frac{dy}{1-y} \ln^2(y)\ln(1+y)
,
\\
\label{hp4-12}
 H[ 0,-1, 1, 1,x] &=&  U_4(x)
- \ln 2 ~H[0,-1,1,x] - \frac{1}{2}\ln^2 2 ~H[0,-1,x],
\\
\label{hp4-13}
  H[ 0, 1,-1, 1,x] &=& - V_4(x)  -2 H[0,-1,1,1,x] 
+ \ln 2~\left(H[1,0,-1,x]-H[1,x]H[0,-1,x]\right), 
\\
\label{hp4-14}
  H[ 0, 1, 1,-1,x] &=& V_4(x) +\frac{1}{2} W_4(x) + H[0,-1,1,1,x]
- \ln 2~\left(H[1,0,-1,x]-H[1,x]H[0,-1,x]\right)
 ,
\\
\label{hp4-15}
 H[ 0, 1,-1,-1,x]  &=& - U_4(-x)
+ \ln2 ~H[0,1,-1,x]
- \frac{1}{2}\ln^2 2~H[0,1,x],
\\
\label{hp4-16}
  H[ 0,-1, 1,-1,x]  &=&  V_{4}(-x) -2 H[0,1,-1,-1,x] 
- \ln 2~\left(H[1,0,-1,x]-H[1,x]H[0,-1,x]\right),
\\
\label{hp4-17}
  H[ 0,-1,-1, 1,x]    &=&  
 - V_4(-x)  
 -\frac{1}{2} W_4(-x)
+ H[0,1,-1,-1,x] 
\nl &&
+~\ln 2~\left(H[1,0,-1,x]-H[1,x]H[0,-1,x]\right)
.
\ea
Further, $H[a,a,a,a,x] = \frac{1}{24} H[a,x]^4$.  
Some auxiliary integrals are defined below in this appendix.
The integrals $K_1$ and $J_2$ in (\ref{hp4-2h2}) 
 may be performed analytically, see  (\ref{hp4-1f}) and
(\ref{hp4-1v}), and that in (\ref{hp4-3}) is not known to us in
analytical form. 
We just mention that the functions with index sets  $(0000)$, $(0001)$,
$(0011)$, $(0111)$, $(1111)$ may also be found in \cite{Gehrmann:2000zt}.
Our  list of HPLs is available in file {\tt HPL4.m} in
\cite{web-masters:2004nn}. 
We give here one explicit example of a HPL of the class (aabc), derived with
equation (4.26) of \cite{Blumlein:2003gb} \footnote{In the relations
  for the harmonic sums given in  \cite{Blumlein:2003gb} one has to
  delete all terms with the operator 
  $\wedge$ in the index set, in order to obtain from the relations for
  harmonic sums the simpler relations for the HPLs.}: 
\ba
\label{hp4-18}
 H[-1, 1, 0, 1,x] &=&  2H[0,1,1,-1,x] + H[0,1,-1,1,x] + H[1,x]( -
 H[1,-1,0,x] - H[1,0,-1,x] 
\nl&&+~ H[-1,0,1,x] + H[0,-1,1,x] ) 
-
 2H[-1,x]H[0,1,1,x] + H[1,-1,x]H[1,0,x].
\ea 
The above list of functions is a quite compact basis which is
sufficient for the computation of all HPLs until weight 4.

For comparisons with \cite{Davydychev:2003mv} we used additionally to
HPLs with arguments $x$ also the following functions:  
\ba
\label{hp4-a1}
\text{Li}_n(x^2) &=& 2^{n-1}\left[ H[0,\ldots,0,1,x] -
  H[0,\ldots,0,-1,x]\right] 
\ea
with (n-1) zeroes in the arguments, and:
\ba
\label{hp4-a2}
S_{1,2}(x^2) &=& H\left[0,1,1,x^2\right] 
= 
2\left[ H[0,1,1,x] - H[0,-1,1,x]- H[0,1,-1,x]+ H[0,-1,-1,x]\right],
\\
\label{hp4-a3}
S_{2,2}(x^2) &=& H\left[0,0,1,1,x^2\right] 
= 
4\left[ H[0,0,1,1,x] - H[0,0,-1,1,x]- H[0,0,1,-1,x]+ H[0,0,-1,-1,x]\right].
\ea
For a study of boundary conditions, one needs the values of HPLs
at $x = 0$ and $x = 1$.
Here we follow the algorithms proposed in \cite{Remiddi:1999ew}. 

We have introduced several auxiliary functions.
For $ H[ 0,-1, 1,x]$ (see (\ref{hp3-int})) we need:
\ba
\label{hp3-212}
V_3(x) 
&=&
\int_0^x\frac{dy}{y}\left[{\litwo}\left(\frac{1+y}{2}\right)-
 {\litwo}\left(\frac{1}{2}\right)\right] 
\nl
&=&
 J_1(x) 
+
\ln(x) 
\left[{\litwo}\left(\frac{1+x}{2}\right)
     -{\litwo}\left(\frac{1}{2}\right)\right]
-~\ln 2 \left[ {\litwo}(-x) + \ln(x) \ln(1+x)\right] ,
\ea
with 
\ba
J_1(x)&=&\int_0^x \frac{dy}{1+y} \ln(y)\ln(1-y)
\nl
&=& 
-{\litr}\left(\frac{1-x}{2}\right) 
-{\litr}\left(-x\right) 
-{\litr}\left(-\frac{x}{1-x}\right) 
+{\litr}\left(-\frac{2x}{1-x}\right) 
\nl&&
+\ln\left(\frac{2x}{1-x} \right) \left[
  {\litwo}\left(-\frac{x}{1-x}\right)
 - {\litwo}\left(-\frac{2x}{1-x}\right) \right]
+\ln\left(2x \right){\litwo}\left(-x\right)
+\ln\left(\frac{1-x}{2} \right){\litwo}\left(\frac{1-x}{2}\right)
\nl&&
+\frac{1}{2} \ln\left(1-x\right) 
\left[-\ln 2 \ln(1-x) + 2 \ln 2 \ln(2x)+ 2
  \ln(x)\ln\left(\frac{1+x}{2}\right) \right] 
-\left(\frac{1}{3}\ln^3 2 - \frac{7}{8}\zeta_3 \right)
.
\ea
The $H[1,1,-1,-1,x]$ (see (\ref{hp4-2h2})) is known analytically, using:
\ba
\label{hp4-1f}
K_1(x) &=& \int_0^x \frac{dy}{1-y}\left[
 S_{1,2}\left(\frac{1-y}{2}\right)
- S_{1,2}\left(\frac{1}{2}\right) \right]  
\nl
&=&
\ln 2 ~J_3(x) - \frac{1}{2} K_2(x)
-\ln(1-x)\left[
 S_{1,2}\left(\frac{1-x}{2}\right)
- S_{1,2}\left(\frac{1}{2}\right) \right] 
+\frac{1}{4} \ln^2 2 \ln^2(1-x) 
,
\\
\label{hp4-1v}
J_2(x) &=&
\int_0^x \frac{dy}{1-y}
\left[{\litwo}\left(\frac{1-y}{2}\right)
  -{\litwo}\left(\frac{1}{2}\right)\right] 
\nl &=&
 J_3(x)
-\ln(1-x)\left[{\litwo}\left(\frac{1-x}{2}\right)
  -{\litwo}\left(\frac{1}{2}\right)\right] 
+\frac{1}{2}\ln 2 \ln^2 (1-x) ,
\ea 
with
\ba
 J_3(x) &=& \int_0^x \frac{dy}{1-y} \ln(1-y) \ln(1+y)
\nl &=&
 - {\litr}\left(\frac{1-x}{2} \right)+
   {\litr}\left(\frac{1}{2} \right) + \ln 
(1-x)\litwo\left(\frac{1-x}{2} \right)  
- \frac{1}{2}\ln 2 \ln^2 (1-x),
\\
 K_2(x) 
&=&  \int_0^x \frac{dy}{1-y} \ln(1-y) \ln^2 (1+y)
\nl
&=&
\frac{7}{10}\zeta_2^2 - \zeta_2\ln^2 2 + \ln^4 2 
+  \frac{1}{4}\ln^4(1+x) 
- \ln^3(1+x)\ln(1-x) 
+ \frac{1}{2} \ln^2(1+x) \ln^2(1-x) 
\nl&&
- \ln 2 \ln(1-x) \ln(1+x) \ln\frac{(1 - x)}{2(1 + x)} 
-  \ln^3 2  ~\ln\left(1 - x^2\right) 
+(\ln^2 2 +\ln^2(1-x))\litwo[(1 - x)/2] 
\nl&&
+\ln^2((1-x)/(1+x))\litwo[-(1 - x)/(1 + x)]
+(\ln^2 2 -\ln^2(1+x))\litwo[(1 + x)/2]
\nl &&
 -2  \ln(1 - x) \litr[(1 - x)/2] 
-2\ln  [(1 - x)/(1 + x)]   \litr [-(1 - x)/(1 + x)] 
\nl &&
+  2 \ln(1 +x)   \litr [(1 + x)/2] 
+ 2\lifo[(1 - x)/2] 
+ 2\lifo[-(1 - x)/(1 + x)] 
- 2\lifo[(1 + x)/2]
.
\ea

In the class $(0,1,1,-1)$ we have an analytical expression for
$(H[0,1,1,-1,x] + H[0,1,-1,1,x])$, but not for $H[0,-1,1,1,x]$ and  
$(H[0,1,1,-1,x] - H[0,1,-1,1,x])$.
An analogous statement applies for the class  $(0,-1,-1,1)$.
The reason is that $W_4(x)$ is known from equations (A.28) to (A.33) of
reference \cite{Davydychev:2003mv}:
\ba
\label{hp4-7d}
W_4(x) &=& 
\int_0^x
 \frac{dy}{y} \ln^2(1- y)\ln(1+ y)
\nl
&=&
-\frac{1}{2} S_{1,3}(x^2) +2S_{1,3}(- x) 
- \left[{\lifo}\left(\omega \right) - {\lifo}\left(-\omega \right) \right]
+ \frac{15}{8}\zeta_4 
\nl &&
+~ \ln(\omega) \left[{\litr}\left(\omega \right) 
  - {\litr}\left(-\omega \right) \right]
-\frac{1}{2}  \ln^2(\omega) \left[{\litwo}\left(\omega \right) 
  - {\litwo}\left(-\omega \right) \right]
-~ \frac{1}{6}  \ln^3(\omega) \ln(x),
\ea
with $\omega= (1-x)/(1+x)$.
By expressing $S_{1,3}(x^2)$ through HPLs with argument $x$
\cite{Remiddi:1999ew}, one may derive the remarkably simple relation
\ba
\label{hp4-7c}
 W_4(x) +  W_4(-x) &=&
- S_{1,3}(x^2) +2  S_{1,3}(x)+2  S_{1,3}(-x).
\ea
This relation follows also from \cite{Davydychev:2003mv}.

We have rewritten the two remaining auxiliary functions:
\ba
\label{hp4-7b}
 U_4(x)
&=&
\int_0^x \frac{dy}{y}\left[S_{1,2}\left(\frac{1+y}{2}\right) -
S_{1,2}\left(\frac{1}{2}\right) \right] 
\nl
&=&
\ln 2 ~J_1(x) 
 -\frac{1}{2} I_3(x)
+
\ln(x)\left[S_{1,2}\left(\frac{1+x}{2}\right) -
S_{1,2}\left(\frac{1}{2}\right) \right] 
\nl&&
-\frac{1}{2}\ln^2 2 \left[\ln(x)\ln(1+x) + \litwo(-x) \right]
,
\\
\label{hp4-7}
V_4(x) &=& 
\int_0^x \frac{dy}{y}\ln(1-y)\left[{\litwo}\left(\frac{1+y}{2}\right)
- {\litwo}\left(\frac{1}{2}\right) \right]
\nl &=&
 - I_2(x)
-\litwo(x) \left[{\litwo}\left(\frac{1+x}{2}\right)
- {\litwo}\left(\frac{1}{2}\right) \right]
\nl &&
+\ln 2 \left[\frac{1}{2}S_{1,2}(x^2)- S_{1,2}(x)- S_{1,2}(-x)
+ \ln(1+x)\litwo(x)   \right] 
.
\ea

We mention finally that we do not know analytical expressions 
for the following auxiliary functions:
\ba
I_1(x)&=& 
\int_0^x \frac{dy}{1+y} {\litr}(y)
,
\\ \label{hp4-7f}
  I_2(x) &=& \int_0^x \frac{dy}{1+y} \litwo(y) \ln(1-y),
\\
\label{hp4-7e}
  I_3(x)&=& \int_0^x \frac{dy}{1+y} \ln(y)\ln^2(1-y),
\\
\label{hp4-7p}
  I_4(x) &=& \int_0^x \frac{dy}{1+y} \ln^2 (y) \ln(1-y).
\ea
The first integral is related to $H[-1,0,0,1,x]$ (see (\ref{hp4-1})), and
the next two integrals to 
four combinations of the six HPLs of classes $(0,a,a,b)$; see
(\ref{hp4-12}) to (\ref{hp4-17}) and the remark related to $W_4$
before (\ref{hp4-7d}).  
The last one is related to $H[-1,1,0,0,x]$, see (\ref{hp4-3}).


\begingroup\begin{thebibliography}{10}

\bibitem{Aguilar-Saavedra:2001rg}
{ECFA/DESY LC Physics Working Group} Collaboration, J.~Aguilar-Saavedra {\em et
  al.}, TESLA TDR, Part III: Physics at an $e^+e^-$ Linear Collider, DESY
  2001-011,
\href{http://www.arXiv.org/abs/hep-ph/0106315}{hep-ph/0106315}.

\bibitem{Hawkings:1999ac}
R.~Hawkings and K.~M{\"o}nig, {\em Eur. Phys. J. direct} {\bf C1} (1999) 8,
\href{http://www.arXiv.org/abs/hep-ex/9910022}{hep-ex/9910022}.

\bibitem{Lohmann:2004nn}
H.~Abramowicz {\em et al.}, {\em IEEE Transactions on Nuclear Science} {\bf 51}
  (2004) 1.

\bibitem{Lohmann:2004nn2}
{FCAL Collaboration, W. Lohmann} {\em et al.},
  http://www-zeuthen.desy.de/TESLA/mask/Collaboration.html.

\bibitem{Jadach:2003zr}
S.~Jadach,
\href{http://www.arXiv.org/abs/hep-ph/0306083}{hep-ph/0306083}.

\bibitem{Jadach:1996is}
S.~Jadach, W.~Placzek, E.~Richter-Was, B.~Ward, and Z.~Was, {\em Comput. Phys.
  Commun.} {\bf 102} (1997)
229--251.

\bibitem{Melles:1997qa}
M.~Melles, {\em Acta Phys. Polon.} {\bf B28} (1997) 1159--1206,
\href{http://arXiv.org/abs/hep-ph/9612348}{hep-ph/9612348}.

\bibitem{Arbuzov:1995qd}
A.~Arbuzov {\em et al.}, {\em Nucl. Phys.} {\bf B485} (1997) 457--502,
\href{http://www.arXiv.org/abs/hep-ph/9512344}{hep-ph/9512344}.

\bibitem{Arbuzov:1996jj}
A.~Arbuzov {\em et al.}, {\em Nucl. Phys. Proc. Suppl.} {\bf 51C} (1996)
  154--163,
\href{http://arXiv.org/abs/hep-ph/9607228}{hep-ph/9607228}.

\bibitem{Glover:2001ev}
N.~Glover, B.~Tausk, and J.~van~der Bij, {\em Phys. Lett.} {\bf B516} (2001)
  33--38,
\href{http://www.arXiv.org/abs/hep-ph/0106052}{hep-ph/0106052}.

\bibitem{Mastrolia:2003yz}
P.~Mastrolia and E.~Remiddi, {\em Nucl. Phys.} {\bf B664} (2003) 341--356,
\href{http://www.arXiv.org/abs/hep-ph/0302162}{hep-ph/0302162}.

\bibitem{Bonciani:2004qt}
R.~Bonciani, A.~Ferroglia, P.~Mastrolia, E.~Remiddi, and J.~van~der Bij,
\href{http://www.arXiv.org/abs/hep-ph/0411321}{hep-ph/0411321}.

\bibitem{web-masters:2004nn}
M.~Czakon, J.~Gluza, and T.~Riemann,
  http://www-zeuthen.desy.de/theory/research/bhabha/.

\bibitem{Passarino:2004nn}
G.~Passarino, {\em Nucl. Phys. Proc. Suppl.} {\bf 135} (2004)
265.

\bibitem{Binoth:2000ps}
T.~Binoth and G.~Heinrich, {\em Nucl. Phys.} {\bf B585} (2000) 741--759,
\href{http://arXiv.org/abs/hep-ph/0004013v.2}{hep-ph/0004013v.2}.

\bibitem{Binoth:2003ak}
T.~Binoth and G.~Heinrich, {\em Nucl. Phys.} {\bf B680} (2004) 375--388,
\href{http://www.arXiv.org/abs/hep-ph/0305234}{hep-ph/0305234}.

\bibitem{Smirnov:2001cm}
V.~Smirnov, {\em Phys. Lett.} {\bf B524} (2002) 129--136,
\href{http://arXiv.org/abs/hep-ph/0111160}{hep-ph/0111160}.

\bibitem{Bonciani:2003cj}
R.~Bonciani, A.~Ferroglia, P.~Mastrolia, E.~Remiddi, and J.~van~der Bij, {\em
  Nucl. Phys.} {\bf B681} (2004) 261--291,
\href{http://www.arXiv.org/abs/hep-ph/0310333}{hep-ph/0310333}.

\bibitem{Heinrich:2004iq}
G.~Heinrich and V.~Smirnov,
\href{http://www.arXiv.org/abs/hep-ph/0406053}{hep-ph/0406053}.

\bibitem{Czakon:2004tg}
M.~Czakon, J.~Gluza, and T.~Riemann, {\em Nucl. Phys. (Proc. Suppl.)} {\bf
  B135} (2004) 83,
\href{http://www.arXiv.org/abs/hep-ph/0406203}{hep-ph/0406203}.

\bibitem{Laporta:1996mq}
S.~Laporta and E.~Remiddi, {\em Phys. Lett.} {\bf B379} (1996) 283--291,
\href{http://www.arXiv.org/abs/hep-ph/9602417}{hep-ph/9602417}.

\bibitem{Laporta:2001dd}
S.~Laporta, {\em Int. J. Mod. Phys.} {\bf A15} (2000) 5087--5159,
\href{http://arXiv.org/abs/hep-ph/0102033}{hep-ph/0102033}.

\bibitem{Remiddi:2004nn}
E.~Remiddi, {\tt SOLVE}, unpublished.

\bibitem{Anastasiou:2004vj}
C.~Anastasiou and A.~Lazopoulos, {\em JHEP} {\bf 07} (2004) 046,
\href{http://www.arXiv.org/abs/hep-ph/0404258}{hep-ph/0404258}.

\bibitem{Czakon:2004uu2}
M.~Czakon, {\tt DiaGen/IdSolver}, unpublished.

\bibitem{Chetyrkin:1981qh}
K.~Chetyrkin and F.~Tkachov, {\em Nucl. Phys.} {\bf B192} (1981)
159--204.

\bibitem{Gehrmann:1999as}
T.~Gehrmann and E.~Remiddi, {\em Nucl. Phys.} {\bf B580} (2000) 485--518,
\href{http://arXiv.org/abs/hep-ph/9912329}{hep-ph/9912329}.

\bibitem{Remiddi:1999ew}
E.~Remiddi and J.~Vermaseren, {\em Int. J. Mod. Phys.} {\bf A15} (2000)
  725--754,
\href{http://www.arXiv.org/abs/hep-ph/9905237}{hep-ph/9905237}.

\bibitem{Kotikov:1991hm}
A.~V. Kotikov, {\em Phys. Lett.} {\bf B259} (1991)
314--322.

\bibitem{Remiddi:1997ny}
E.~Remiddi, {\em Nuovo Cim.} {\bf A110} (1997) 1435--1452,
\href{http://www.arXiv.org/abs/hep-th/9711188}{hep-th/9711188}.

\bibitem{Bonciani:2003hc}
R.~Bonciani, P.~Mastrolia, and E.~Remiddi, {\em Nucl. Phys.} {\bf B690} (2004)
  138--176,
\href{http://www.arXiv.org/abs/hep-ph/0311145}{hep-ph/0311145}.

\bibitem{Bonciani:2003te}
R.~Bonciani, P.~Mastrolia, and E.~Remiddi, {\em Nucl. Phys.} {\bf B661} (2003)
  289--343,
\href{http://www.arXiv.org/abs/hep-ph/0301170}{hep-ph/0301170}.

\bibitem{Consoli:1979xw}
M.~Consoli, {\em Nucl. Phys.} {\bf B160} (1979)
208.

\bibitem{Bohm:1984yt}
M.~B{\"o}hm, A.~Denner, W.~Hollik, and R.~Sommer, {\em Phys. Lett.} {\bf B144}
  (1984)
414.

\bibitem{Bohm:1986fg}
M.~B{\"o}hm, A.~Denner, and W.~Hollik, {\em Nucl. Phys.} {\bf B304} (1988)
687.

\bibitem{Bardin:1990xe}
D.~Bardin, W.~Hollik, and T.~Riemann, {\em Z. Phys.} {\bf C49} (1991)
485--490.

\bibitem{Bardin:1997xq}
D.~Bardin {\em et al.},
\href{http://www.arXiv.org/abs/hep-ph/9709229}{hep-ph/9709229}.

\bibitem{Beenakker:1997fi}
W.~Beenakker and G.~Passarino, {\em Phys. Lett.} {\bf B425} (1998) 199--207,
\href{http://www.arXiv.org/abs/hep-ph/9710376}{hep-ph/9710376}.

\bibitem{Bardin:1999yd}
D.~Bardin {\em et al.}, {\em Comput. Phys. Commun.} {\bf 133} (2001) 229--395,
\href{http://www.arXiv.org/abs/hep-ph/9908433}{hep-ph/9908433}.

\bibitem{Kobel:2000aw}
{Two Fermion Working Group} Collaboration, M.~Kobel {\em et al.},
\href{http://www.arXiv.org/abs/hep-ph/0007180}{hep-ph/0007180}.

\bibitem{Gluza:2004tq}
J.~Gluza, A.~Lorca, and T.~Riemann, {\em Nucl. Instrum. Meth.} {\bf 534} (2004)
  289,
\href{http://www.arXiv.org/abs/hep-ph/0409011}{hep-ph/0409011}.

\bibitem{Lorca:2004dk}
A.~Lorca and T.~Riemann, {\em Nucl. Phys. Proc. Suppl.} {\bf 135} (2004) 328,
\href{http://www.arXiv.org/abs/hep-ph/0407149}{hep-ph/0407149}.

\bibitem{Lorca:2004fg}
A.~Lorca and T.~Riemann,
\href{http://www.arXiv.org/abs/hep-ph/0412047}{hep-ph/0412047}.

\bibitem{'tHooft:1972fi}
G.~'t~Hooft and M.~Veltman, {\em Nucl. Phys.} {\bf B44} (1972)
189--213.

\bibitem{Surguladze:1989ez}
L.~Surguladze and F.~Tkachov, {\em Comput. Phys. Commun.} {\bf 55} (1989)
205--215.

\bibitem{Bonciani:2001}
R.~Bonciani.
\newblock PhD thesis, University of Bologna, 2001.

\bibitem{Fleischer:1999tu}
J.~Fleischer and M.~Kalmykov, {\em Comput. Phys. Commun.} {\bf 128} (2000)
  531--549,
\href{http://arXiv.org/abs/hep-ph/9907431}{hep-ph/9907431}.

\bibitem{Broadhurst:1990ei}
D.~Broadhurst, {\em Z. Phys.} {\bf C47} (1990)
115--124.

\bibitem{Broadhurst:1991fi}
D.~Broadhurst, {\em Z. Phys.} {\bf C54} (1992)
599--606.

\bibitem{Fleischer:1998nb}
J.~Fleischer, A.~Kotikov, and O.~Veretin, {\em Nucl. Phys.} {\bf B547} (1999)
  343--374,
\href{http://arXiv.org/abs/hep-ph/9808242}{hep-ph/9808242}.

\bibitem{Davydychev:2003mv}
A.~I. Davydychev and M.~Y. Kalmykov, {\em Nucl. Phys.} {\bf B699} (2004) 3--64,
\href{http://www.arXiv.org/abs/hep-th/0303162}{hep-th/0303162}.

\bibitem{Fleischer:1999hp}
J.~Fleischer, M.~Kalmykov and A.~Kotikov,
 {\em Phys.\ Lett.}  {\bf B462} (1999) 169,
 \href{http://www.arXiv.org/abs/hep-ph/9905249}{hep-ph/9905249}. 


\bibitem{Czakon:2004wu}
M.~Czakon, J.~Gluza, and T.~Riemann,
\href{http://www.arXiv.org/abs/hep-ph/0409017}{hep-ph/0409017}.

\bibitem{Boos:1990rg}
E.~Boos and A.~Davydychev, {\em Theor. Math. Phys.} {\bf 89} (1991)
1052--1063.

\bibitem{Smirnov:2004ip}
V.~Smirnov,
\href{http://www.arXiv.org/abs/hep-ph/0406052}{hep-ph/0406052}.

\bibitem{Smirnov:book4}
V.~Smirnov, {\em Evaluating Feynman Integrals}, vol.~211 of {\em Springer
  Tracts in Modern Physics}.
\newblock Springer, Berlin, 2004.

\bibitem{Vermaseren:1998uu}
J.~Vermaseren, {\em Int. J. Mod. Phys.} {\bf A14} (1999) 2037--2076,
\href{http://arXiv.org/abs/hep-ph/9806280}{hep-ph/9806280}.

\bibitem{Moch:2001zr}
S.~Moch, P.~Uwer, and S.~Weinzierl, {\em J. Math. Phys.} {\bf 43} (2002)
  3363--3386,
\href{http://www.arXiv.org/abs/hep-ph/0110083}{hep-ph/0110083}.

\bibitem{Moch:2002hm}
S.~Moch, P.~Uwer, and S.~Weinzierl, {\em Phys. Rev.} {\bf D66} (2002) 114001,
\href{http://www.arXiv.org/abs/hep-ph/0207043}{hep-ph/0207043}.

\bibitem{Gehrmann:2000zt}
T.~Gehrmann and E.~Remiddi, {\em Nucl. Phys.} {\bf B601} (2001) 248--286,
\href{http://www.arXiv.org/abs/hep-ph/0008287}{hep-ph/0008287}.

\bibitem{Argeri:2002wz}
M.~Argeri, P.~Mastrolia, and E.~Remiddi, {\em Nucl. Phys.} {\bf B631} (2002)
  388--400,
\href{http://arXiv.org/abs/hep-ph/0202123}{hep-ph/0202123}.

\bibitem{Nakanishi:1971}
N.~Nakanishi, {\em Graph Theory and {Feynman} Integrals}.
\newblock Gordon and Breach, 1971.

\bibitem{Hepp:1966eg}
K.~Hepp, {\em Commun. Math. Phys.} {\bf 2} (1966)
301--326.

\bibitem{Roth:1996pd}
M.~Roth and A.~Denner, {\em Nucl. Phys.} {\bf B479} (1996) 495--514,
\href{http://www.arXiv.org/abs/hep-ph/9605420}{hep-ph/9605420}.

\bibitem{Denner:2004iz}
A.~Denner and S.~Pozzorini,
\href{http://www.arXiv.org/abs/hep-ph/0408068}{hep-ph/0408068}.


\bibitem{Czakon:2004ww}
M.~Czakon, {\tt sectors.m}, unpublished.

\bibitem{Hahn:2004fe}
T.~Hahn,
\href{http://www.arXiv.org/abs/hep-ph/0404043}{hep-ph/0404043}.

\bibitem{Blumlein:2003gb}
J.~Bl{\"u}mlein, {\em Comput. Phys. Commun.} {\bf 159} (2004) 19--54,
\href{http://www.arXiv.org/abs/hep-ph/0311046}{hep-ph/0311046}.

\bibitem{Moch:1999eb}
S.~Moch and J.~Vermaseren, {\em Nucl. Phys.} {\bf B573} (2000) 853--907,
\href{http://www.arXiv.org/abs/hep-ph/9912355}{hep-ph/9912355}.

\bibitem{Bern:2000ie}
Z.~Bern, L.~Dixon, and A.~Ghinculov, {\em Phys. Rev.} {\bf D63} (2001) 053007,
\href{http://www.arXiv.org/abs/hep-ph/0010075}{hep-ph/0010075}.

\bibitem{Czakon:2002wm}
M.~Czakon, J.~Gluza, and J.~Hejczyk, {\em Nucl. Phys.} {\bf B642} (2002)
  157--172,
\href{http://www.arXiv.org/abs/hep-ph/0205303}{hep-ph/0205303}.

\bibitem{Awramik:2002wn}
M.~Awramik and M.~Czakon, {\em Phys. Rev. Lett.} {\bf 89} (2002) 241801,
\href{http://www.arXiv.org/abs/hep-ph/0208113}{hep-ph/0208113}.

\bibitem{Czakon:2004bu}
M.~Czakon,
\href{http://www.arXiv.org/abs/hep-ph/0411261}{hep-ph/0411261}.

\bibitem{Lewis:2004nn}
R.~H. Lewis, Fermat, http://www.bway.net/\verb+~+lewis/.

\bibitem{Vermaseren:2000nd}
J.~Vermaseren,
\href{http://www.arXiv.org/abs/math-ph/0010025}{math-ph/0010025}.

\bibitem{Hahn:1998yk}
T.~Hahn and M.~P{\'e}rez-Victoria, {\em Comput. Phys. Commun.} {\bf 118} (1999)
  153,
\href{http://arXiv.org/abs/hep-ph/9807565}{hep-ph/9807565}.

\bibitem{Hahn:2001xx}
T.~Hahn, ``{LoopTools User's Guide}''. http://www.feynarts.de/looptools/.

\end{thebibliography}\endgroup

\providecommand{\href}[2]{#2}

\end{document}